# Persistent Homology generalizations for Social Media Network Analysis

Isabela ROCHA[1]
Institute of Political Science
of the University of Brasília

**Abstract:** This study details an approach for the analysis of social media collected data through the lens of Topological Data Analysis (TDA), with a specific focus on Persistent Homology and the political processes they represent by proposing a set of mathematical generalizations using density functions to define and analyze these Persistent Homology categories. In the context of an overarching research on Digital Social Media and the 2022 Brazilian online election campaign involving the scraping and analysis of a 2 million tweet database, three distinct types of Persistent Homologies were recurrent across datasets that had been plotted through retweeting patterns and analyzed through the k-Nearest-Neighbor (kNN) filtrations. As these Persistent Homologies continued to appear, they were then categorized and dubbed Nuclear, Bipolar, and Multipolar Constellations. Upon investigating the content of these plotted tweets, specific patterns of interaction and political information dissemination were identified, namely Political Personalism and Political Polarization. Through clustering and application of Gaussian density functions, I have mathematically characterized each category, encapsulating their distinctive topological features: the Nuclear Constellation by a singular, dense focal point, the Bipolar Constellation by two prominent clusters, and the Multipolar Constellation by several dispersed clusters. The mathematical generalizations of Bipolar, Nuclear, and Multipolar Constellations developed in this study are designed to inspire other political science digital media researchers to utilize these categories as to identify Persistent Homology in datasets derived from various social media platforms, suggesting the broader hypothesis that such structures are bound to be present on political scraped data regardless of the social media it's derived from. By providing a clear, topologically grounded framework for categorizing data structures, this method aims to offer a new perspective in Network Analysis as it allows for a nuanced exploration of the underlying shape of the networks formed by retweeting patterns, enhancing the understanding of digital interactions within the sphere of Computational Social Sciences. This methodology extrapolates Network Analysis by leveraging the nuanced capabilities of TDA, offering a refined understanding of complex social media data structures.

---

[1] Master of Science and PhD Candidate at the Institute of Political Science of the University of Brasília (IPOL-UnB). Coordinator of the Working Group on Strategy, Data and Sovereignty of the Study and Research Group on International Security of the Institute of Foreign Affairs of the University of Brasília (GEPSI IREL-UnB). Member of the Public Information and Elections Research Group at the Institute of Political Science of the University of Brasília (IPê UnB) and the International Political Science Association 50th Research Committee on the Politics of Language (IPSA RC50).

**Introduction**

The advent of Digital Social Media has revolutionized the way information is created, shared, and consumed, providing a continuous stream of user-generated data that is unparalleled in both volume and accessibility. Unlike traditional methods of data collection, which often require extensive resources to encourage active participation, social media platforms naturally elicit a vast array of content from users, ranging from opinions and personal experiences to direct responses to political events and policies. This spontaneous and organic generation of data presents an invaluable opportunity for political scientists, specially during periods of political significance such as elections, offering a real-time, unfiltered window into the public sphere. Thus, the field of Computational Social Sciences has re-emerged as a pivotal domain, bridging advanced computational techniques with the intricate dynamics of human behavior and societal structures as individuals willingly produce massive amounts of data.

Such sheer volume and complexity of available data through digital social media, despite its spontaneous nature, also present unique challenges. Extracting meaningful insights from millions of tweets, posts, and interactions requires not just computational power, but also a nuanced understanding of the underlying patterns and structures within this data. Although Network Analysis methods have already provided a compelling toolbox for researchers, the sheer volume and complexity of data being produced necessitates further exploration into more sophisticated methodologies, enhancing our ability to make scientifically robust assumptions in the face of multi-dimensional and dynamically evolving datasets. Addressing these challenges, this paper introduces a methodological approach that harnesses the power of Topological Data Analysis (TDA), with a special focus on Persistent Homology, to analyze and categorize patterns of interaction and political information dissemination. This approach allows us to navigate the vast sea of data with an eye for the underlying topological structures, providing a fresh perspective in the analysis of digital interactions and their implications in the political landscape.

This article's research findings are derived from overarching research on Politics on Digital Social Media with an emphasis on processes of Political Personalism and Political Polarization, and discovered upon analyzing a 2 million tweet database collected throughout the 2022 Brazilian elections. Upon plotting the collected data through traditional Network Analysis methods, three generalized shapes were recurrent: **Nuclear, Bipolar** and **Multipolar Constellations**. The tweets were scraped from political trending topics during the official electoral campaign, and then plotted into networks by the retweeting patterns through the use of the "RT @user tweet-string" interface on Twitter's 1.1 API endpoint. As the similar shapes continued to appear across different datasets, a research question emerged: is there a correlation between political processes and geometrical shapes?

The recurrence of these Persistent Homology categories in the datasets points to an inherent structure within the retweet networks on Twitter, underscoring the value of this analytical approach. By adopting a topological lens, this study not only enriches the traditional methods of Network Analysis but also opens new avenues for understanding the nuances and complexities of digital social interactions. In short, the identification of these homologies provides a more granular understanding of the social media landscape, offering insights into the dynamics of information dissemination, community formation, and the evolving nature of digital communication.

Upon discovering a correlation between Nuclear Constellations and processes of Political Personalism, and Bipolar Constellations and processes of Political Polarization, in twitter-scraped data, Persistent Homology generalizations were formulated, in hopes of provoking other researchers on finding such shapes in their own Social Media Network Analysis plotted data. This article presents such generalizations for each of the Persistent Homologies, a sample of generalized data which is then plotted and filtered through the k-Nearest-Neighbor application (kNN), and, of course, real data that fit into the proposed categories, also filtered in kNN as to display the persistence of the homologies, as proposed by Le and Taylor (2022). Here, I also address the limitations of Gaussian-based generalization models, as symmetries are observed and data is fluid and continually evolving. While the data samples used for analysis are static, representing specific snapshots in time, the underlying data is dynamic, reflecting the constantly changing landscape of social media discourse. This dynamic nature challenges the static models and necessitates ongoing adaptation of analytical frameworks to capture the real-time evolution of social interactions on digital platforms.

**Persistent Homology generalizations through k-Nearest-Neighbors and the Political Processes they represent**

In the dynamic landscape of social media analytics, the application of traditional Network Analysis methods has consistently provided valuable insights into patterns of digital communication. This study leveraged such methods to analyze data obtained via the 'RT @user tweet content' interface of Twitter's API 1.1, allowing for Network Graphs to be constructed through a Wolfram Language code dubbed Argos. As the data was collected, over the course of the 2022 Brazilian electoral campaign, and plotted, distinct and recurrent shapes emerged: Nuclear, Bipolar and Multipolar.

Such recurrence called for further investigation, and as to analyze the underlying structure of the data, the Networks were plotted into a dimensional space, and then investigated through the kNN filtrations to reveal Persistent Homologies, as suggested by Le and Taylor (2022): In their study, they develop kNN complexes as an extension of kNN graphs, leading to the formulation of kNN-based Persistent Homology techniques. Le and Taylor's approach is particularly relevant in scenarios where the relative positioning of data points is more important than their precise locations – such as Network Graphs built from retweeting patterns, and this aspect makes kNN-based Persistent Homology a valuable tool in various data science applications, and, of course, most suitable for Social Media scraped data.

The choice to employ the k-Nearest-Neighbor (kNN) algorithm in this study is grounded in its intuitive approach and its established utility in Network Analysis methodologies, employed on this research as user Networks were plotted through their retweets. Notably, this approach is adopted even though the networks under examination were derived from retweeting patterns, as the overarching research hoped to capture the aspect of social interaction and information dissemination on the platform. Retweeting patterns inherently form a network structure based on user interactions and shared content, but the use of kNN in this context is particularly advantageous due to its ability to intuitively capture these relational dynamics by inferring neighborhoods based on the correlations. In other words, while the raw data from Twitter's API outlines a clear connection map based on retweets, kNN allows us to delve deeper, identifying subtler structures and relationships within this network. By analyzing the nearest neighbors in the dataset, kNN filtration helps to uncover the underlying topological structures that are not immediately apparent in the straightforward retweet networks. In summary, the application of kNN to retweeting pattern data, despite its straightforward initial network structure, offers a more comprehensive understanding of the Persistent Homologies identified. This methodology aligns with the objectives of this study, highlighting kNN's invaluable role in mapping and interpreting the intricate topologies of digital social landscapes.

The structure of the shapes only became clearer as the kNN filtrations were applied. The recurrent shapes of Nuclear, Bipolar, and Multipolar Constellations were analytically explored using these kNN filtrations, thereby unveiling the topological structures that underlie social media interactions and discourse and raising the question: is there a connection between the observed Persistent Homology, and Social Media political processes? Nuclear Constellations were observed in scenarios where tweets clustered around a central, dominant theme or opinion, indicative of a focused dialogue or a unified response to a specific event or topic. Bipolar Constellations, on the other hand, manifested in contexts characterized by a clear division of opinions, reflecting the polarized nature of public discourse, particularly prevalent in political discussions. Finally, Multipolar Constellations emerged in more complex and diverse dialogues, where multiple groups or opinions were interacting simultaneously, showcasing the pluralistic nature of social media conversations.

It is also important to mention the reasoning behind the choice of kNN filtrations as opposed to the most traditional and stablished Vietoris-Rips (VR) filtrations, which is driven by the nature of the used data and the specific analytical objectives. Social Media data, inherently relational and network-based, presents a unique landscape where the relative positioning of data points (i.e., user interactions) often holds more analytical value than their exact geometric placement – even though the networks had to be plotted into a bidimensional space to enable the kNN filtrations. kNN is particularly suited for this scenario because it focuses on the relationships and proximities between points, making it ideal for interpreting and mapping the complex dynamics of virtual networks like those formed through retweeting patterns on social media platforms. In contrast, the VR complex is more traditionally used in contexts where the precise geometric or spatial relationships between data points are of primary interest. It is highly effective in physical or 'real-world' scenarios where the exact distances between points are crucial for understanding the underlying structure of the data. For example, in fields like sensor networks, biological data analysis, or astronomy, the Vietoris-Rips approach is invaluable for constructing topologies based on fixed distance thresholds.

When it comes to the fluid and dynamic realms of social media, where the emphasis is on the interaction patterns and information flow, the kNN approach offers a more nuanced and relevant analysis. By considering the nearest neighbors in a social network, kNN filtration can reveal hidden structures and relationships that go beyond mere connectivity, highlighting clusters, influential nodes, and subgroups that might be obscured in a traditional VR analysis (Le, Taylor, 2022). This ability to tease out the more subtle and contextually relevant relational dynamics makes kNN a more suitable tool for the analysis of social media data, providing insights into how users cluster around topics, ideologies, and influencers in the digital space.

Finally, another matter must be brought to attention before we venture into the Constellations: the necessary data reduction due to limited RAM capacity, which impacts the distribution and appearance of both randomly generated and real data sets. For the application of the kNN algorithm, any sample larger than a little more than 3000 points will not be processed, thanks to the hardware at my disposal. The necessary reduction process, essential for managing limited computational resources, has influenced how the real data aligns with the Gaussian generalization used in our analyses. For the randomly generated data intended for displaying the generalization, starting with a reduced sample leads to a more disperse distribution, lessening the density but preserving the general Gaussian distribution's characteristics. Such a reduction method maintains the overall shape and spread of the Gaussian function, which can still adequately represent the underlying distribution patterns despite a visually sparser appearance, but, when compared to the real data, seems less dense.

In the case of the real data, which originates from network-based structures, the reduction was implemented through a Euclidean-based method to begin with, to preserve the overarching geometrical shapes by maintaining distances and spatial relationships as typically measured in a Euclidean space. This approach aims to maintain the spatial relationships and configurations crucial for accurate topological analysis, such as those used in Persistent Homology. However, this method can result in a less dense nucleus in the reduced dataset compared to the full data – which is the framework of a Gaussian function. The lesser density in the nucleus arises because the reduction method selectively retains points that maintain the overall shape, which may include removing some points from denser areas. For example, I have employed the Euclidean-based reduction into a sample of randomly generated coordinates from the Gaussian equation framework. In this case, the nucleus seems less dense, even if the data remains nucleated.

**Figure 1** – Side by side comparison of a 34000 randomly generated sample of coordinates and its Euclidean-based reduction. Beneath, a sample of 3400 randomly generated coordinates.

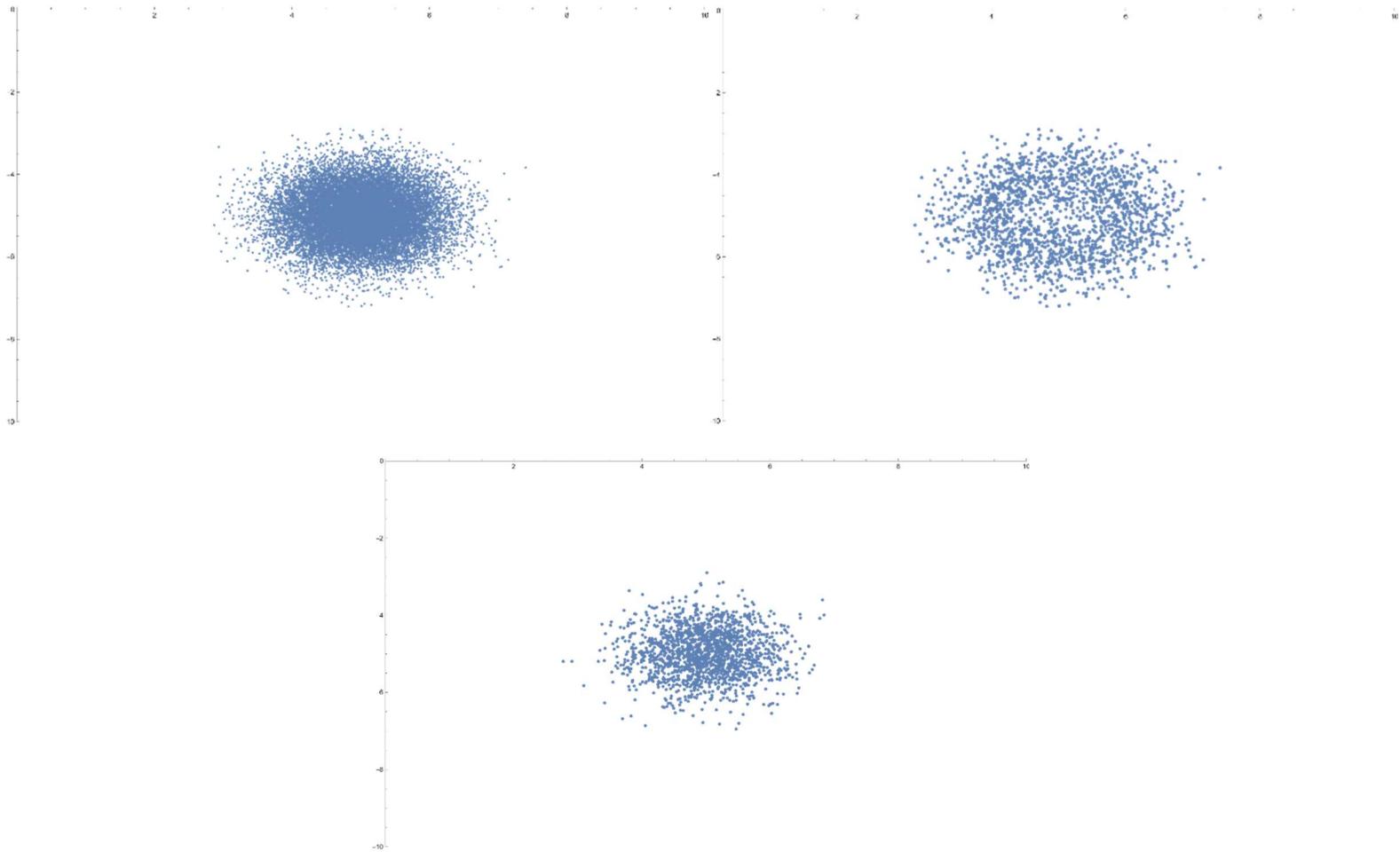

Source: Author.

An obvious issue of this method of simplification is the appearance of a subtle hole in the structure of the reduced data, as seen in Figure 1. This problem tends to be more pronounced with randomly generated data, where the reduction process might inadvertently emphasize artificial gaps that do not correspond to any meaningful features in the data. In contrast, with real-world data, although similar gaps – holes – might also appear, they are generally not as significant or evident unless they correspond to actual features of the underlying topology. In other words, holes in real data are much more pronounced than the subtle holes that appear when the Euclidean-based reduction is employed.

The presence of holes, however, in real data, typically offers valuable insights into the dataset's intrinsic structure and is often a crucial element in topological analysis, such as Persistent Homology. Unfortunately, the constraints imposed by limited RAM and processing capabilities mean that some level of data reduction is inevitable. This necessitates a careful balance between data fidelity and computational feasibility, ensuring that the most important topological features are preserved even when working within hardware limitations. The next figures showcases a sample of plotted data that *actually* displays a hole and its Euclidean-based reduction.

**Figure 2** – Real plotted data displaying a clear hole and a kNN filtration of Euclidean-based reduced data where k = 3. As opposed to Figure 1, the hole here is a feature in itself.

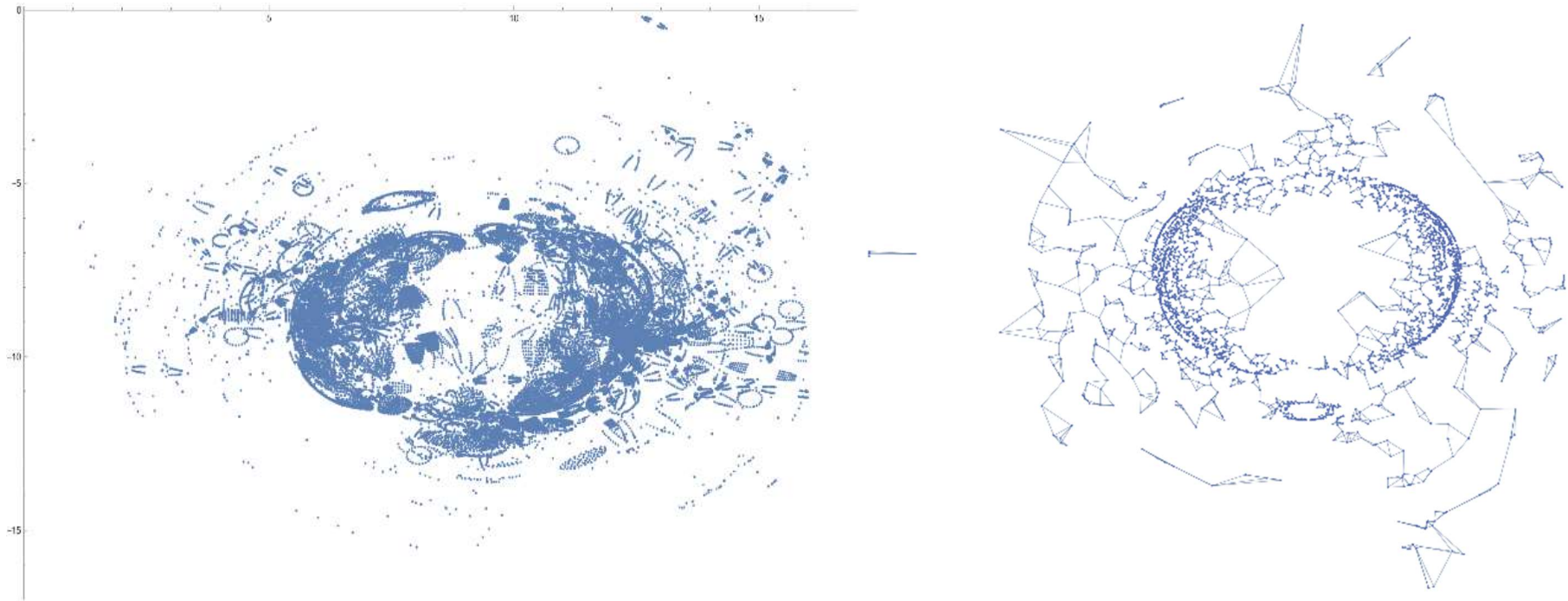

Source: Author.

Despite the implications of the data reduction, it remains feasible to fit the reduced datasets into the Gaussian generalization framework for several reasons. Firstly, the core characteristics of the Gaussian model, which fundamentally describes how data points are distributed around a central point within a specific distance determined by the standard deviation $\sigma$, are preserved. The reduction process may alter point density but generally does not impact this central relationship key for categorizing the Persistent Homology used to define the Constellations, especially when it maintains the shape and relative distances through methods like Euclidean-based reduction. Additionally, the parameters of the Gaussian distribution, particularly $\sigma$, can be adjusted to reflect changes in scale and dispersion that result from data reduction. This allows the model to still provide a good fit and robust generalization, even if the reduced data appears less dense or more spread out. Moreover, Persistent Homology, which focuses on the shapes and features that persist across scales, helps ensure that the essential topological features the Gaussian model seeks to represent are retained. By preserving the overall shape and connectivity of the data through the reduction, significant topological features that inform the Gaussian generalization are effectively maintained.

Let us, then, look into each category, from their Generalizations to Real Data and the qualitative information one may extract:

**Nuclear Constellations**

Nuclear Constellations in the context of social media data analysis are characterized by a dense, central cluster of data points, indicating a focused area of interaction or discussion. This dense core, surrounded by less densely populated areas, suggests a significant concentration of activity or attention around a specific topic, individual profile, or event, which led me to explore Gaussian equations for the generalizations. Such relation between Nuclear Constellations and Gaussian equations lies in the way these constellations could be mathematically modeled: A Gaussian distribution is often used to describe the spread of points around a central core, much like a Nuclear Constellation. In this model, the peak of the Gaussian distribution represents the dense central point, and the spread of the distribution indicates the point density decrease as they move away from the center.

The Gaussian function used in this context may be expressed as:

$$f(x,y) = exp\left(-\frac{(x-x_c)^2 + (y-y_c)^2}{2\sigma^2}\right)$$

Here, $(x_c, y_c)$ represents the coordinates of the centroid of the Nuclear Constellation, $\sigma$ denotes the standard deviation reflecting the spread or dispersion of data points around the centroid, and $(x, y)$ are the coordinates of any point in the constellation. This Gaussian function effectively captures the essence of a Nuclear Constellation by modeling the high concentration of data points at the center and the gradual decrease in density outward from the center.

The complete generalization may, then, be written as:

$$Nuclear\ Constellation = \{C, R, f(x,y)\}$$

$$f(x,y) = exp\left(-\frac{(x-x_c)^2 + (y-y_c)^2}{2\sigma^2}\right)$$

The centroid, denoted as $C$, represents the pivotal point around which the data points in the constellation are clustered. In social media data analysis, this centroid could be a particular topic, event, or individual that garners significant focus and engagement. It's the epicenter of discussion or activity, around which conversations, or interactions densely aggregate. Mathematically, the centroid is represented by coordinates, typically expressed as $(x_c, y_c)$, serving as a quantitative anchor for the constellation's location within the data space.

The radius, denoted as $R$, quantifies the spread or extent of the constellation around its centroid as a measure of the reach or influence of the central user or theme within the dataset. Essentially, $R$ can be understood as the average distance from the centroid to the surrounding points, defining the boundary within which the core activity or discussion is concentrated. This radius helps in delineating the area of influence, providing insights into how far the impact of the central theme extends in the social media landscape.

Finally, the function $f(x,y)$ describes the density of data points around the centroid, as previously described. This Gaussian function, expressed as $f(x,y) = exp\left(-\frac{(x-x_c)^2+(y-y_c)^2}{2\sigma^2}\right)$, where $\sigma$ is the standard deviation that indicates the dispersion or spread of data points around the centroid to determine the clustering pattern and intensity within the constellation.

To display the function, I have generated a sample of coordinates that suit the Nuclear Constellation Generalization, and applied upon them the kNN algorithm to showcase the Persistent Homology that makes a Nuclear Constellation:

**Figure 3 -** A set of 34000 random coordinates generated from the Nuclear Constellation Generalization parameters and its Euclidean-based reduction

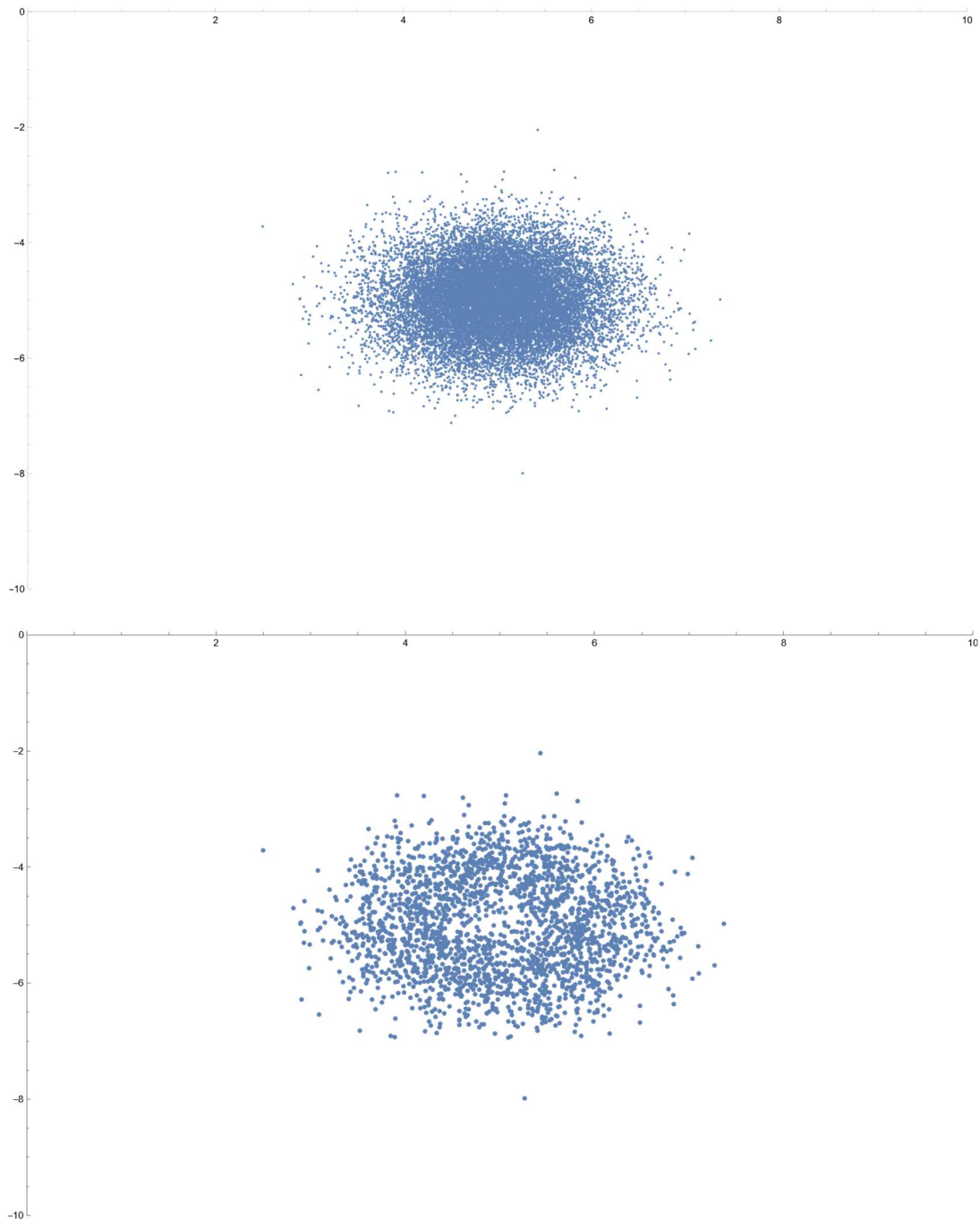

Source: Author

**Figure 4** – kNN filtrations applied upon the reduced sample to showcase the Persistent Homology.

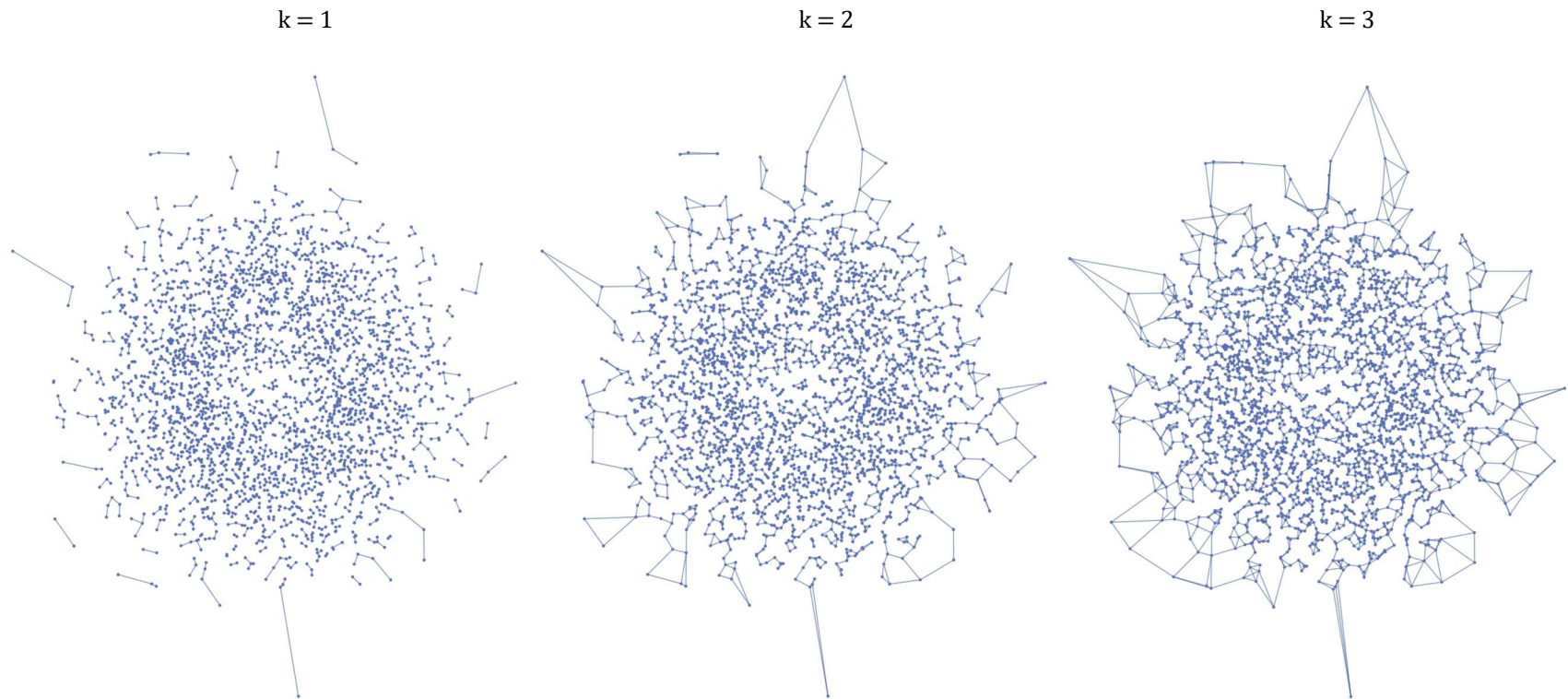

Source: Author.

*Real Data*

Nuclear Constellations might be the most common Persistent Homologies discovered in the context of this research. This prevalence aligns with the observed trend in social media behavior, where individuals often engage with like-minded users. Social media platforms, by their design, foster environments where users are more likely to encounter and interact with content that resonates with their existing beliefs and preferences. This phenomenon, sometimes referred to as the 'echo chamber' effect (Sunstein, 2017), is exemplified in the Nuclear Constellation model, where a central idea or opinion forms the nucleus around which interactions cluster. In these constellations, the centroid represents a dominant viewpoint, drawing in users who share similar perspectives. The discussions within these clusters tend to reinforce existing opinions, as the algorithmic curation of content on many social media platforms amplifies this effect by prioritizing content that aligns with the users' previous interactions. This leads to a concentration of discussion that is not only focused on a particular topic or opinion but also homogenous in its viewpoint, reflecting a narrower slice of the broader public discourse.

In these constellations, the focused clustering around key political figures highlighted the central role of individual personalities in shaping public discourse and political alignment. This phenomenon is particularly pronounced in the electoral context, where personal charisma, reputation, and public perception of candidates can heavily influence voter behavior and political dialogue. On Twitter, this pattern of Nuclear Constellations is not exclusive to politics but would likely be seen in various domains, from entertainment to public health. However, the impact of these constellations is especially profound in the electoral context due to the high stakes involved: Political Personalism, where the focus is more on individual leaders than on policies or party ideologies, can significantly affect the democratic process (Silva, 2022). It often leads to a polarization of opinions, where discussions are less about debating ideas and more about supporting or opposing specific individuals.

It has become evident that Bolsonaro's digital presence is predominantly bolstered by right-wing influencers[2] who align themselves with his political ideology as

---

[2] It was decided to exempt the username of "person" users to preserve their privacy. The criteria for considering a user a person, as opposed to an influencer is: the use of a proper name (such as Maria da Silva, José de Souza, Fabio Carvalho, etc.) as a username along with a personal photo on the profile and

well as his sons and their respective supporters. These influencers, through their substantial following and persuasive communication strategies, amplify Bolsonaro's message and mobilize support from like-minded individuals. In 2018, social media platforms such as Twitter and WhatsApp played a key role in Jair Bolsonaro's electoral campaign, and although it would not be fully appropriate attribute his victory to social media alone, Bolsonaro himself attributed his election to his online presence, having developed strategies to engage with the media and utilizing his son Flávio's municipal candidacy as a laboratory to develop models, test communication strategies, profile potential voters, and craft narratives that could resonate with potential supporters (Santini, et al., 2021). Bolsonaro recognized the power of social media platforms, particularly Twitter, as an effective tool for engaging with his base and disseminating his political worldview. By leveraging the support of right-wing influencers who share his views, Bolsonaro was able to create a digital network of supporters who actively promoted his message and amplified his reach.

The following constellation (Figure 3) consists of tweets collected on August 16[th] – the very first day of official campaigning as ordained by Brazil's Supreme Electoral Tribunal (STF) – containing one of Bolsonaro's first campaign hashtags: *#Vote22Bolsonaro*. One may realize, even without the kNN filtrations, that these profiles are tightly knit into a single cluster, with only a few retweets disconnected from that main cluster. This makes sense, of course, within the scope of a campaign hashtag, as these will most often be populated by like-minded individuals who will more likely interact with one another. The following user Network was built from a sample of 18000 tweets collected from the *#Vote22Bolsonaro* hashtag, with tweets collected on August 16[th] in the spam of little more than 3 hours. The total community consists of 7456 profiles. The ten most retweeted profiles are labeled.

---

fewer than 100,000 followers. Any named profile with a self-picture over 100,000 followers will be treated as an influencer profile, and their username will be disclosed in this article. These are called "person profiles".

**Figure 5 -** The Network Graph built from the retweeting user patterns of tweets containing the hashtag *#Vote22Bolsonaro*

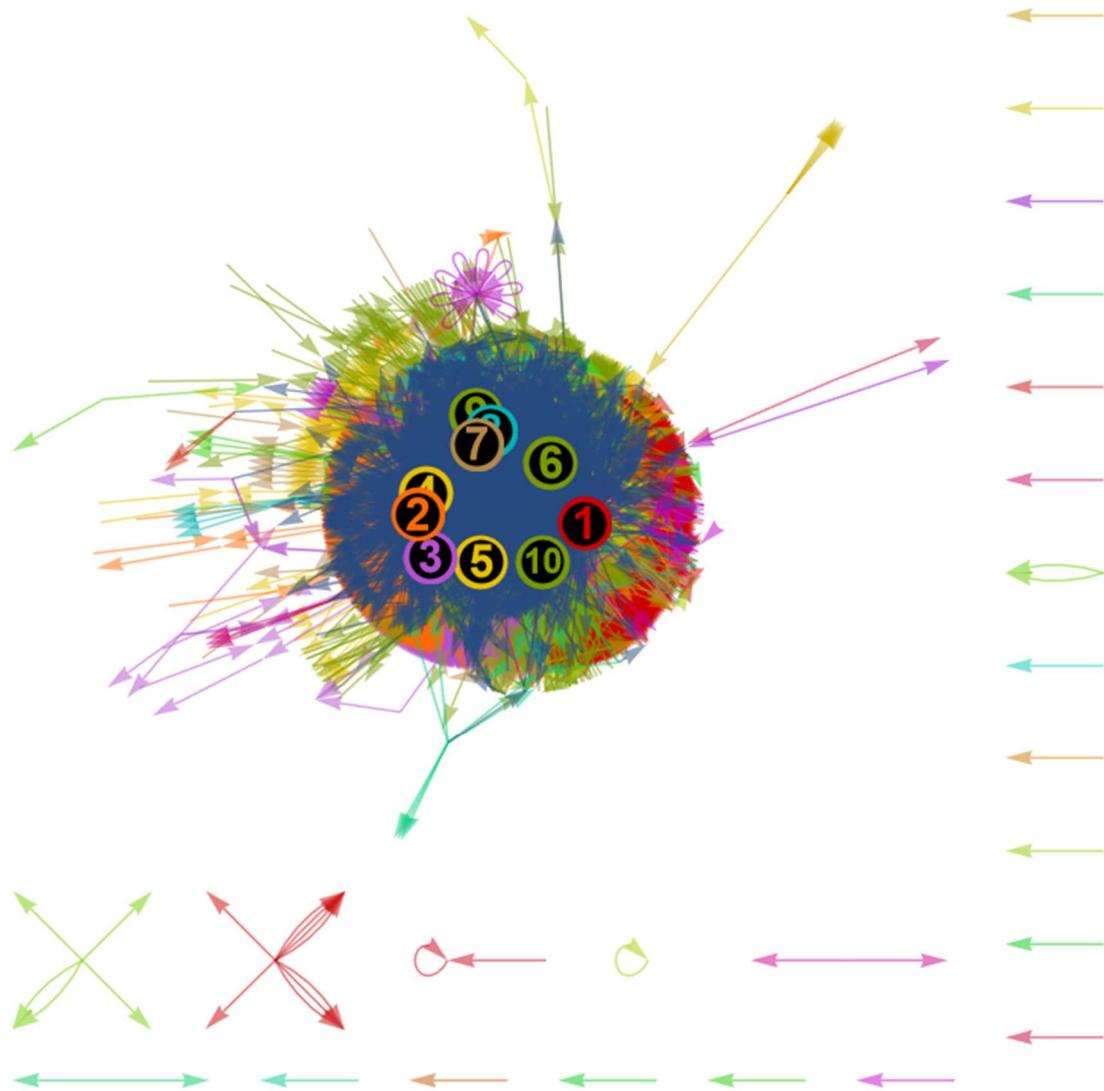

Source: Author.

As expected, the most retweeted profile labeled in red is that of Flávio (@FlavioBolsonaro), Bolsonaro's eldest son, followed by Jouberth Souza (@Jouberth19), candidate for Congress and member of the right-wing party PL (*Partido Liberal*, the Liberal Party). On his tweets, Flavio mentions the assassination attempt Bolsonaro suffered in 2018 at Juiz de Fora, location he, Jouberth and others, were present at the time of data collection. The fourth profile labeled in yellow belongs to the Influencer Bárbara Destefani, known by her profile name TeAtualizei (which roughly translates to "*Kept you up to date*") or her @taoquei1[3] username, and she will often appear amongst Bolsonaro's most influential supporters. The following profile, labeled in yellow and the number 5, is also an influencer: Freu Rodrigues (@freu_rodrigues), though not nearly as influential as Bárbara. The seventh profile is that of Carlos Bolsonaro (@CarlosBolsonaro), Jair Bolsonaro's second-born son, followed by Rodrigo Constantino (@RConstantino), an influential writer and political commentator, and, within Twitter's sphere, an Influencer by all means and purposes. Remarkably, all other profiles (the third, the sixth, and the seventh to the tenth profile) are person profiles.

---
[3] "*Tá ok?*" is a slang often used by Bolsonaro which colloquially translates to "All good?", which, in Portuguese, sounds like "*taoquei*".

A narrative may be widespread from a single event that triggers a viral response, leading into major shifts in politics and voter behavior (Shiller, 2019). The attempt on President Jair Bolsonaro's life during the 2018 presidential campaign was a significant event that lifted his figure to that of legend – and the topic surrounding the displayed Network. On September 6, 2018, Bolsonaro was stabbed during a campaign rally in Juiz de Fora, Minas Gerais, a shocking act of violence that aimed to eliminate Bolsonaro as a political contender. In the aftermath of the attack, his supporters rallied around him, viewing the assassination attempt as evidence of the systemic corruption and political turmoil that needed to be addressed, and, whether the attempt on Bolsonaro's life was decisive for his victory on the 2018's elections or not, it is without doubt that this event built much of the *ethos* surrounding Bolsonaro's persona. His full name is, after all, Jair *Messias* (Messiah, in Portuguese) Bolsonaro. This incident and his survival ultimately played a significant role in solidifying Bolsonaro's image as a strongman leader and thus suit him as a personalist figure.

The official campaign for Jair Bolsonaro's re-election in 2022 began in Juiz de Fora, the same city where he had survived the assassination attempt four years earlier. This strategic decision aimed to capitalize on the symbolic significance of such event: By launching his campaign in Juiz de Fora, Bolsonaro sought to reinforce the narrative of resilience and strength that had emerged from the 2018 attack, presenting himself as a leader who had overcome adversity and was unwavering in his commitment to fighting corruption and restoring order, despite any attempts to his life. Let us now investigate the underlying topological structures present on this Network:

**Figure 6** – The coordinates selected from the *#Vote22Bolsonaro* main Network and it's Euclidean-based reduction.

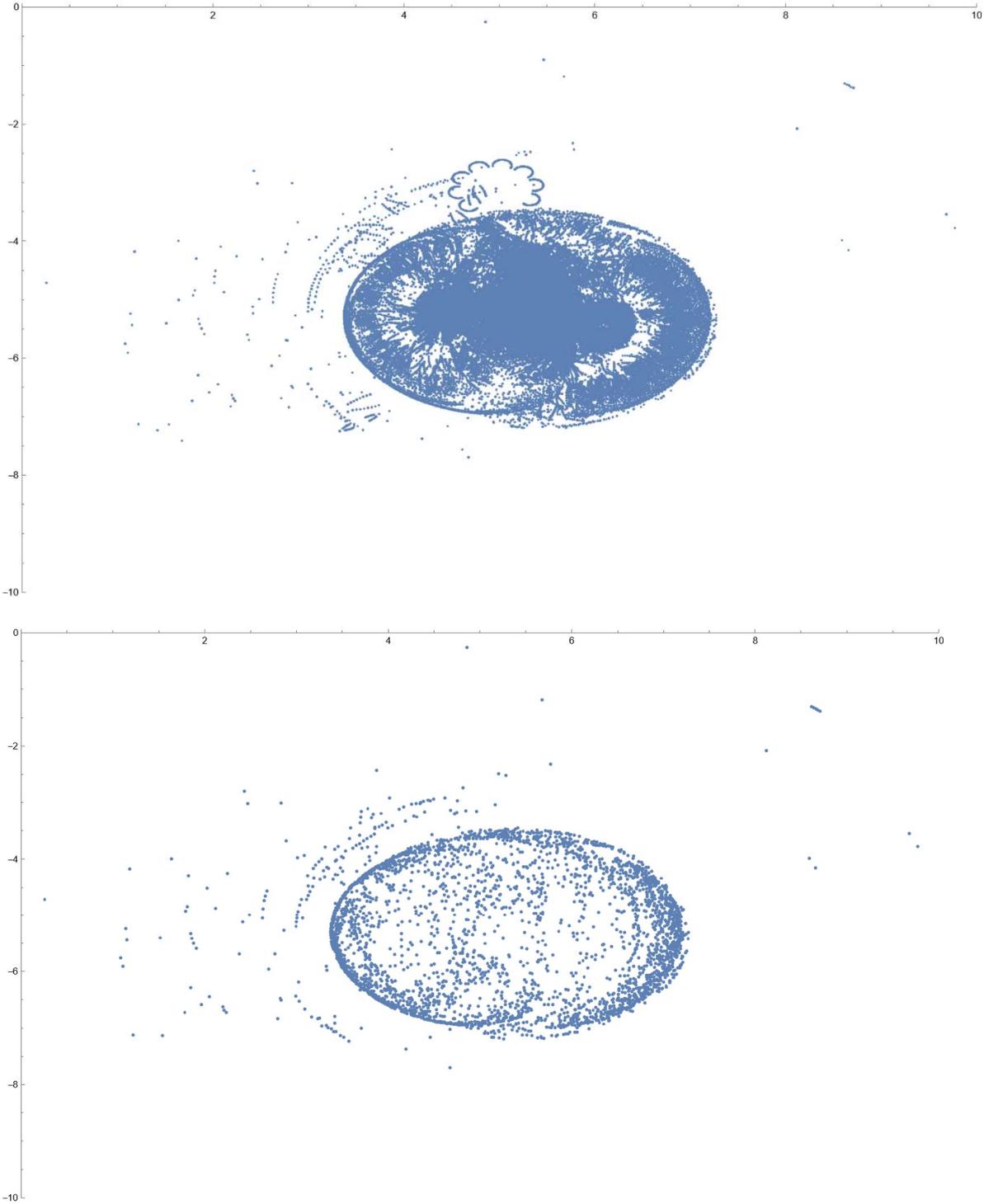

Source: Author.

**Figure 7** – kNN filtrations applied upon a reduced sample[4] of the #Vote22Bolsonaro network coordinates to showcase the Persistent Homology.

k = 1            k = 2            k = 3

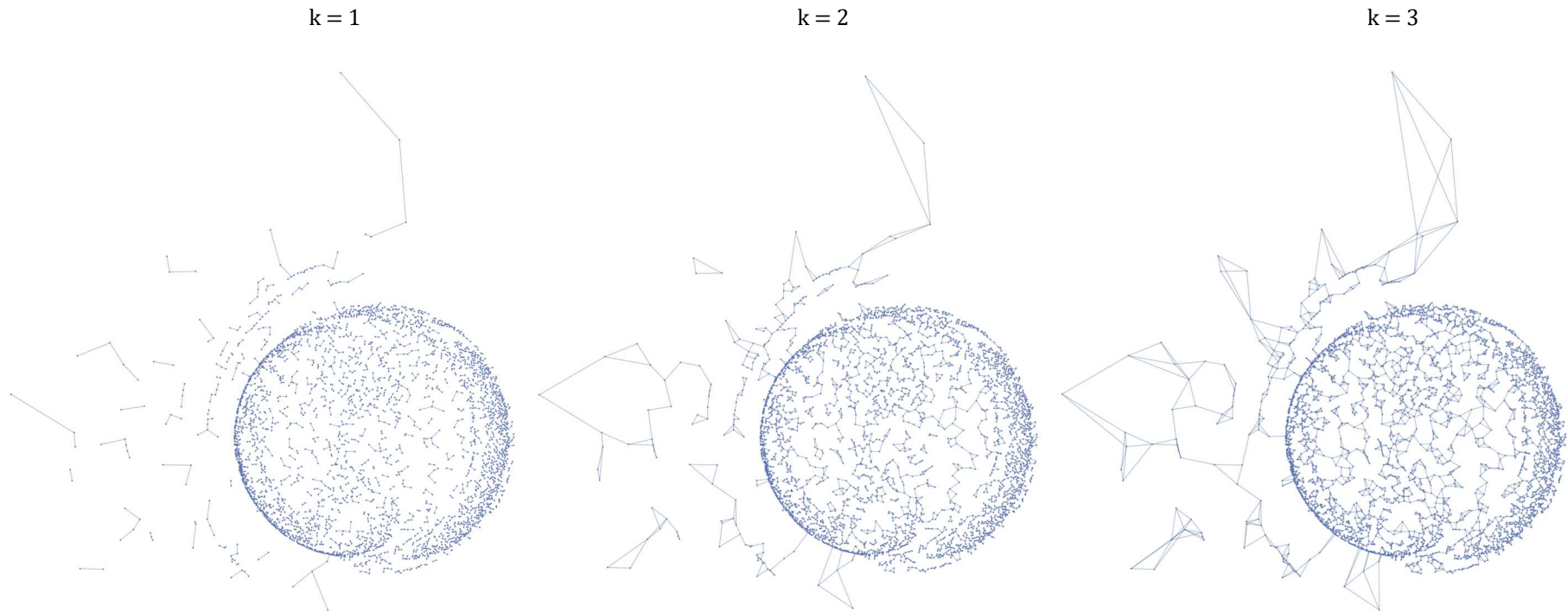

Source: Author.

---

[4] The number of points used for this kNN filtration had to be reduced from the original 69621 points to 3415 to accommodate limited computation power.

The topological structure of this network, underscored by the patterns of clustering, aligns with the quintessential characteristics of a Nuclear Constellation as designed by a single Gaussian function. At the core of this network, the most retweeted profiles, notably that of Flávio Bolsonaro and key influencers such as Jouberth Souza, form a centralized structure. This centrality is emblematic of Nuclear Constellations, where discussions and interactions are densely clustered around a central theme or, in this case, personalities whose discourse is mutually aligned. In this instance, the network is centered around prominent political figures, reinforcing the model where central nodes significantly influences the network's overall discourse and orientation.

This network also exhibits a distinct echo chamber effect, an aspect synonymous with Nuclear Constellations. Such echo chambers are especially prevalent in the realms of political hashtags, which naturally attract like-minded individuals. Consequently, the discussions within this network predominantly reflect a homogenous viewpoint, concentrating on specific perspectives and narratives. This effect is magnified by the use of campaign hashtags, which inherently bring together individuals with similar political alignments, thereby consolidating the dialogue around a unified narrative.

Finally, the topology revealed through this analysis is significantly influenced by the concept of political personalism: Figures like Flávio Bolsonaro and other influencers dominate the discourse while bolstering the pro-Bolsonaro discourse, illustrating the predominance of personalism in digital political engagement. This focus on individual personalities, as opposed to broader political ideologies or policies, characterizes the Nuclear Constellation pattern observed in the network. The centrality of these individuals in the network underscores their impact on shaping and directing the social media narrative. Additionally, the strategic narrative construction, particularly the symbolic significance of Bolsonaro's campaign launch in Juiz de Fora, exemplifies how certain events and narratives can define and influence the topology of a social media network. This event, with its historical and emotional resonance, forms a pivotal point around which the network's discussions revolve, further cementing the Nuclear Constellation structure. This is true not to this data sample alone, but is present across datasets on the broader database of the 2022 Brazilian elections, having been selected for examination in the scope of this article as a good example due to its clear and distinct illustration of the Nuclear Constellation phenomenon, with topologies and Persistent Homologies that are quite similar to the generalizations.

**Figure 8** – Generalization sample (34000 points generated from the gaussian function, and then reduced by their Euclidean distance) and real data kNN (k = 3) comparison.

Generalization                                   Real reduced data

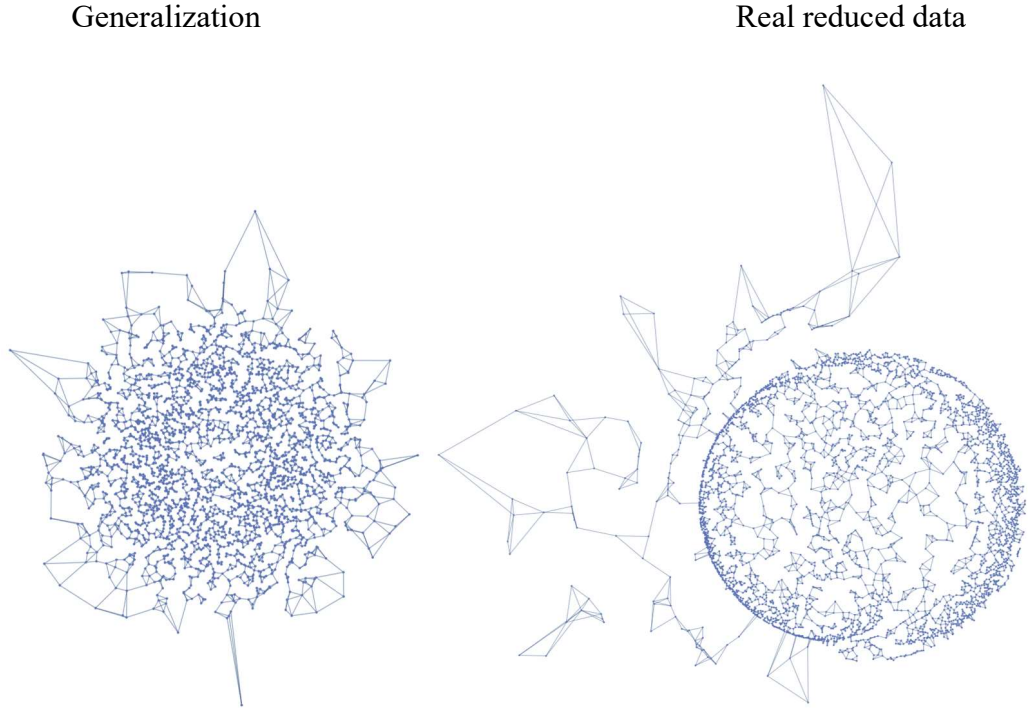

Source: Author

**Bipolar Constellations**

Bipolar Constellations, as suggested by the name, indicate processes of Political Polarization in Digital Social Media. They are characterized by two dense clusters of data points, indicating opposing areas of interaction or discussion. These dense cores, surrounded by less densely populated areas and few interactions between the nearer peripheries of the clusters, might also be defined by Gaussian functions:

$$\textbf{\textit{Bipolar Constellation}} = \{\boldsymbol{C_1, C_2, D, f_1(x,y), f_2(x,y)}\}$$

$$\boldsymbol{f_1(x,y) = exp\left(-\frac{(x-x_1)^2 + (y-y_1)^2}{2\sigma_1^2}\right)}$$

$$\boldsymbol{f_2(x,y) = exp\left(-\frac{(x-x_2)^2 + (y-y_2)^2}{2\sigma_2^2}\right)}$$

$$\boldsymbol{D = \sqrt{(x_2-x_1)^2 + (y_2-y_1)^2}}$$

Here, $(x_1, y_1)$ and $(x_2, y_2)$ represent the coordinates of the centroids of the two dense clusters within the Bipolar Constellation. These centroids serve as the epicenters for each cluster, symbolizing the focal points of two distinct groups, ideologies, or discussions. The Gaussian functions $f_1(x, y)$ and $f_2(x, y)$ describe the distribution of data points around each centroid, modeling the concentration and spread of activity within each cluster.

The parameter $\sigma_1$ in $f_1(x, y)$ and $\sigma_2$ in $f_2(x, y)$ denote standard deviations for each cluster, reflecting how tightly or loosely the data points are grouped around their respective centroids. A smaller $\sigma$ value indicates a more focused and intense engagement around the centroid, while a larger $\sigma$ suggests a more dispersed interaction pattern.

Finally, the distance $D$ between the centroids, calculated as $D = \sqrt{(x_2 - x_1)^2 + (y_2 - y_1)^2}$, is critical in understanding the degree of separation or opposition between the two clusters. This metric can provide quantitative information regarding the polarization levels within the dataset. A larger distance suggests a greater divide, highlighting significant opposition and less interactions between the clusters. This is particularly relevant in social media contexts where opposing groups might form around different opinions, political beliefs, or social issues. Moreover, analyzing the interaction patterns or lack thereof between the peripheries of these clusters can offer additional insights into the nature of the discourse. For instance, if there are sparse interactions between the edges of the clusters, it might indicate a strong barrier to communication and understanding between the groups. Conversely, more interactions might suggest areas where the opposing views are closer to reconciliation or debate. Looking into the profiles, or topics that provide the connective tissue between these opposing clusters can further enrich our understanding of the dynamics at play. Here, the role of gatekeepers becomes pivotal. These individuals or entities serve as critical nodes within the network, not just contributing to the narrative but actively shaping the flow and focus of discussions, able to either mitigate or intensify the process of polarization.

To display the function, I have generated a sample of coordinates that suit the Bipolar Constellation Generalization, and applied upon them the kNN algorithm on its reduced sample to showcase the Persistent Homology that makes a Nuclear Constellation:

**Figure 9** – A set of 34000 coordinates generated from the Bipolar Constellation Generalization parameters and it's Euclidean-based reduction.

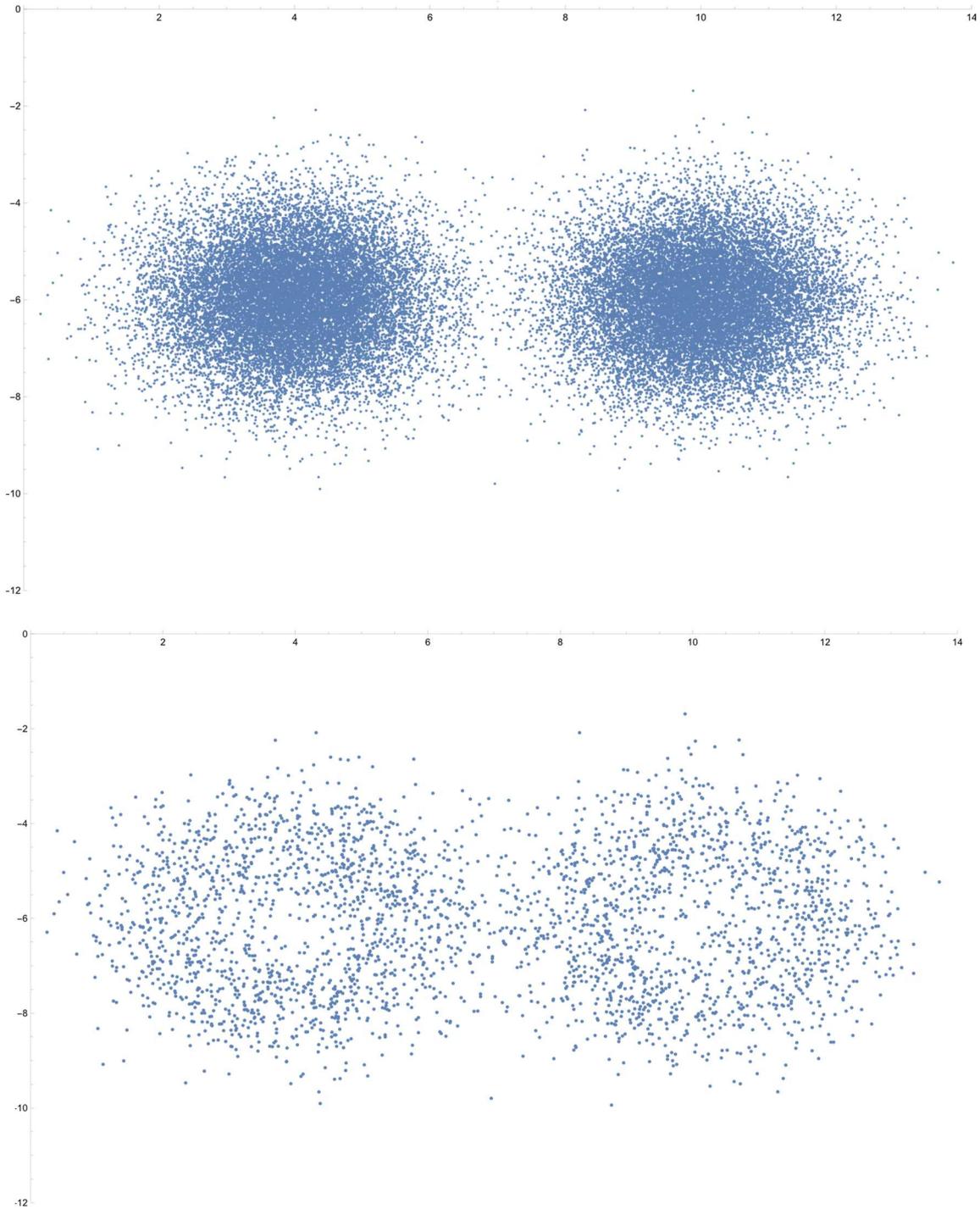

Source: Author.

**Figure 10** – kNN filtrations applied upon the reduced sample to showcase the Persistent Homology.

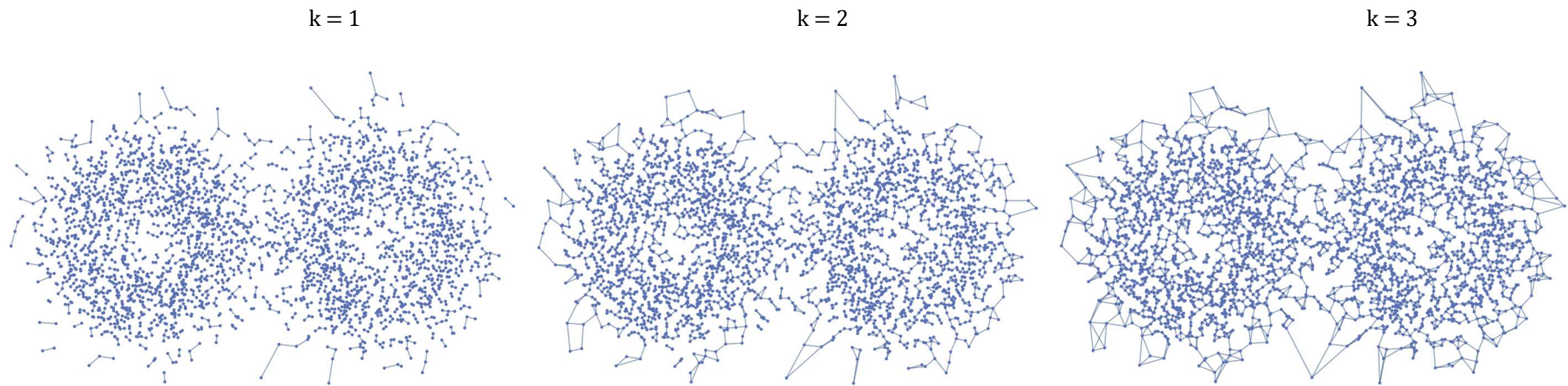

Source: Author.

*Real Data*

Bipolar Constellations have, across datasets within the Brazilian 2022 elections database, displayed a clear competition between two generally homogeneous opinion groups. This pattern was particularly evident in the context of Brazilian politics, where the analysis revealed a distinct dichotomy in social media discourse. The datasets whose networks formed a Bipolar topology predominantly showcased two polarized opinion clusters: one pro-Lula and anti-Bolsonaro, and the other pro-Bolsonaro and anti-Lula, underscoring a significant polarization within the public opinion, as expressed on social media platforms. These Bipolar Constellations not only reflected divergent political stances but also demonstrated a high degree of internal consistency within each group; in the sense that the pro-Lula faction consistently shared content supportive of Luiz Inácio Lula da Silva, while expressing opposition to Jair Bolsonaro, and vice versa. For this paper, I have selected a 18000 tweet dataset built from tweets containing the 'Datafolha' keyword:

Datafolha is a renowned Brazilian polling institute widely recognized for its extensive research and surveys on public opinion. With a solid reputation for accuracy and reliability, Datafolha conducts polls and gathers data on various socio-political issues, including elections, social trends, and public sentiment. Its findings and analyses are often cited by media outlets, politicians, and researchers, contributing to informed discussions and shaping the understanding of Brazilian society. Back in August 18th, it predicted Lula's victory with 51% of the votes, which was nearly correct: Lula won with 50,83% of the votes against Bolsonaro's 49,17%.

**Figure 11** – The Network Graph built from the retweeting user patterns of tweets containing the word 'Datafolha'.

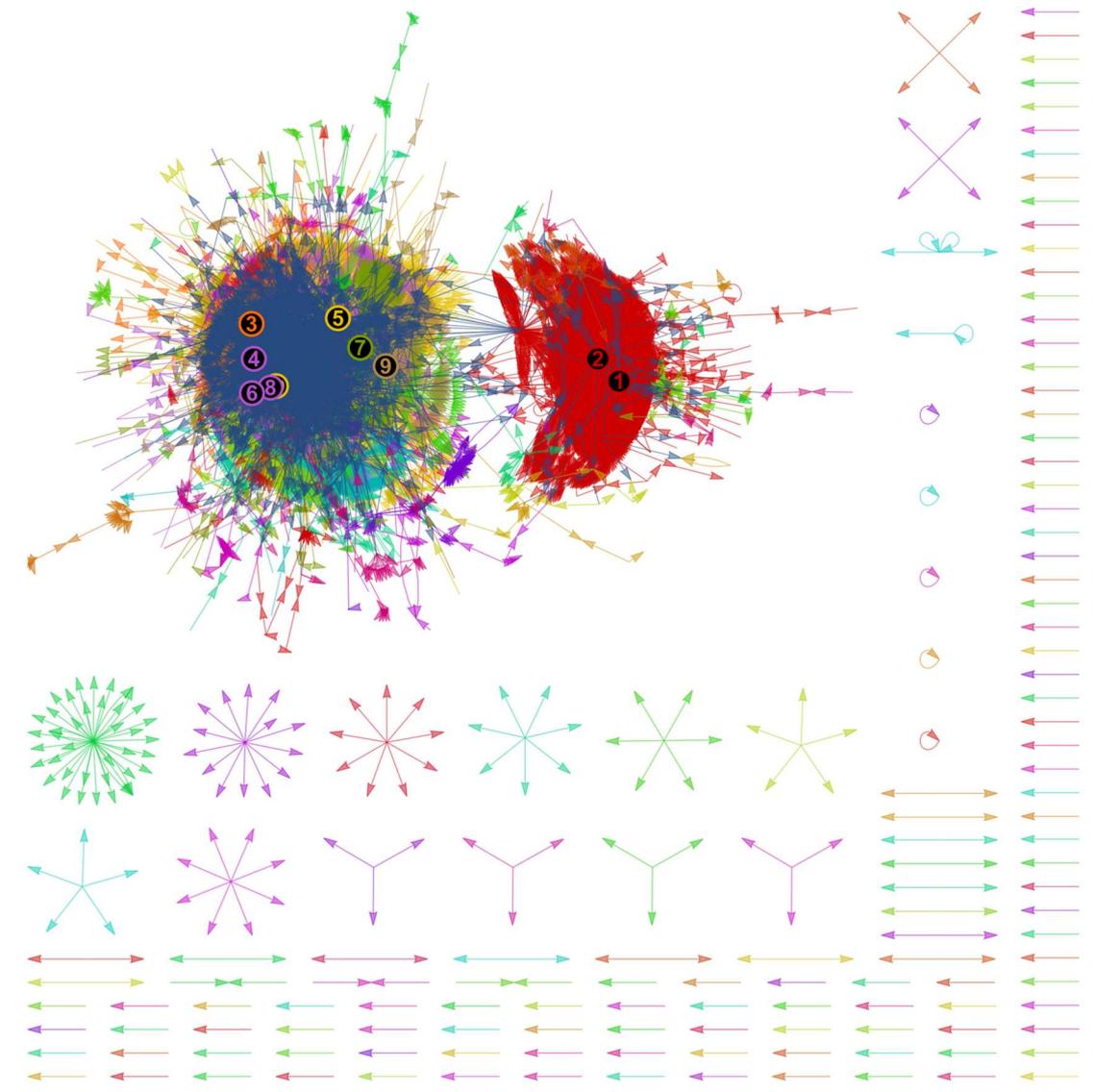

Source: Author.

The most retweeted profile is Flavia Ferronato's (@flferronato), in red, criticizing IPEC[5] survey showing Lula with 51% of voting intention, joking that the only "51[6]" Lula has is in the Brazilian brand of cachaça, implying him as a drunkard. Kim D. Paim (@kimpaim), also in red, emerges as the second most retweeted profile, raising criticism regarding the surveys and highlighting that Datafolha featured smiling photos of all candidates except Bolsonaro, creating a perception that he appeared less sympathetic. He also points out that Globo rectified the situation by changing Bolsonaro's photo, addressing the initial discrepancy and implying that Globo was aware of their bias.

---

[5] Another institute that also conducts researches and surveys on public opinion and polls, often confused with Datafolha.
[6] 51 Cachaça is a popular brand of Brazilian distilled spirit, commonly known as cachaça. It is a widely recognized and consumed alcoholic beverage in Brazil, often used for making cocktails such as the famous Caipirinha.

This time, Argos places the political left on the left. André Janones (@AndreJanonesAdv) and Thiago Brasil (@ThiagoResiste), labeled in orange and purple respectively, are left-wing profiles, sharing the good news and bolstering Lula, followed by Central Eleitoral (@CentralEleicoes) the fifth most retweeted profile in yellow, a left-wing news network that holds influence across various left-leaning communities. Then, Boulos (@GuilhermeBoulos) the sixth most retweeted profile in purple, another politician, supports Lula, followed by two very interesting profiles: Jairme (@jairmearrependi, which loosely translates to "Jair and I regretted it) and Coronel Siqueira (@direitasiqueira), respectively the seventh and eight most retweeted profiles of the sample in green and purple. Jairme is a profile that compilates a plethora of tweets, posts or other samples of posts from those who regretted voting for Bolsonaro, and Coronel (Colonel) Siqueira is a satire profile, where he pretends to be a middle-aged military man who supports Bolsonaro – of course, mockingly so.

One aspect of this search highlights an advantage of the use of TDA: by effectively capturing the underlying topology of the data through the retweeting patterns, identifying the relationships between profiles, it correctly positions Coronel Siqueira's satire profile among the left-leaning profiles, leading to an accurate representation of the ideological clusters within the dataset.

The 'Datafolha' constellation reveals the hole between profiles as well as their positionings, as seen in Figure 10 reveal a clear dispute of narratives surrounding the credibility of polling institutes, particularly Datafolha, and the portrayal of political candidates. Profiles aligned with the right cluster, such as Flavia Ferronato and Kim D. Paim, express skepticism towards Datafolha's survey results, questioning its accuracy and highlighting potential biases. Their criticisms focus on perceived discrepancies in the treatment of candidates, such as the portrayal of Bolsonaro in comparison to other contenders. On the other hand, left-wing profiles like André Janones, Thiago Brasil, and Central Eleitoral celebrate and amplify the positive findings for Lula, bolstering his image and generating enthusiasm among their respective communities. Additionally, the presence of profiles like Jairme, dedicated to showcasing regrets from former Bolsonaro supporters, and Coronel Siqueira, a satirical account that mocks Bolsonaro's stance, further demonstrates the existence of conflicting narratives within the political discourse. The dispute of narratives in this context reflects differing interpretations of polling data, candidate representation, and ideological perspectives, contributing to a complex and dynamic political landscape.

Let us, now, explore its topological features:

**Figure 12 -** The coordinates selected from the *'Datafolha'* main Network and it's Euclidean-based reduction.

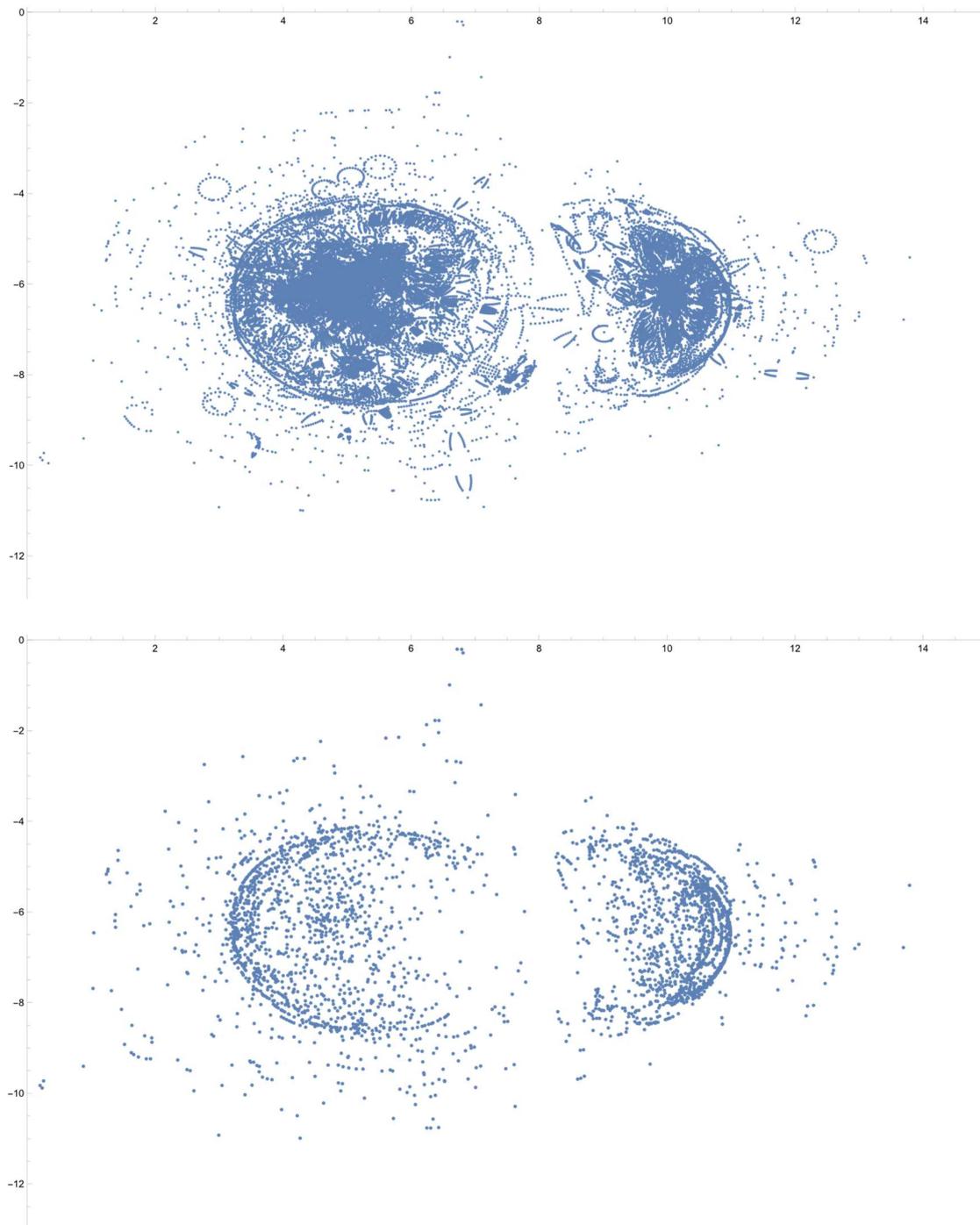

Source: Author

**Figure 13** – kNN filtrations applied upon a reduced sample[7] of the '*Datafolha*' network coordinates to showcase the Persistent Homology.

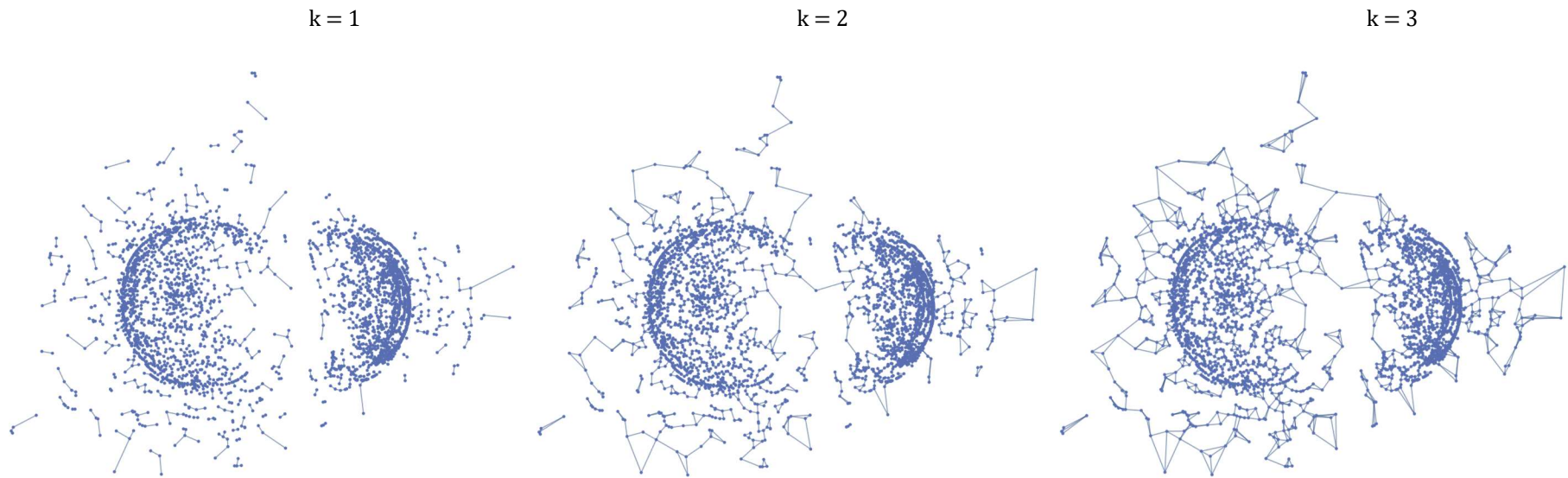

Source: Author.

---

[7] The number of points used for this kNN filtration had to be reduced from the original 37445 points to 3001 to accommodate limited computation power.

In the context of a Bipolar Constellation, there is a stark division between two primary ideological groups, represented by two distinct clusters within the network graph, each aligning closely with either pro-Lula or anti-Bolsonaro sentiments or vice versa, demonstrating a clear polarization within public opinion as expressed on social media platforms.

The centroids of these clusters are represented by the most influential profiles within each group. For the right-leaning cluster, Flavia Ferronato and Kim D. Paim act as central figures around whom discussions and retweeting patterns densely congregate. Their roles are pivotal in disseminating narratives that critique polling methodologies and media biases, particularly against Bolsonaro. Conversely, the left-leaning cluster orbits around profiles like André Janones and Thiago Brasil, who amplify positive polling results for Lula and share content that supports his candidacy. These profiles serve as hubs in their cluster, generating and spreading content that aligns with pro-Lula sentiments.

The distance between these clusters, both ideologically and in terms of the interaction patterns on Twitter, highlights the degree of polarization. There is minimal interaction between the clusters, indicating a strong separation and lack of dialogue between the opposing groups. This separation is further emphasized by the ideological content shared within each cluster, which predominantly reinforces the internal consensus and rarely engages with the opposing viewpoints. In this sense, each cluster acts as an echo chamber, where the shared content and discussions largely reinforce the existing beliefs of the group members. The retweeting patterns emphasize the circulation of messages within the same ideological group, with little crossover to the opposing side. This phenomenon intensifies the polarization, as each group becomes more entrenched in its viewpoints. The narrative disputes are particularly evident in the criticisms levied by right-leaning profiles against the polling methods and media representations, suggesting bias and manipulation. Meanwhile, the left-leaning profiles use the same polling data to bolster their confidence in Lula's electoral prospects, showcasing a classic example of how the same information can be interpreted in starkly different ways depending on ideological leanings.

**Figure 14** – Generalization sample (34000 points generated from the gaussian function, and then reduced by their Euclidean distance) and real data kNN (k = 3) comparison.

Generalization                    Real reduced data

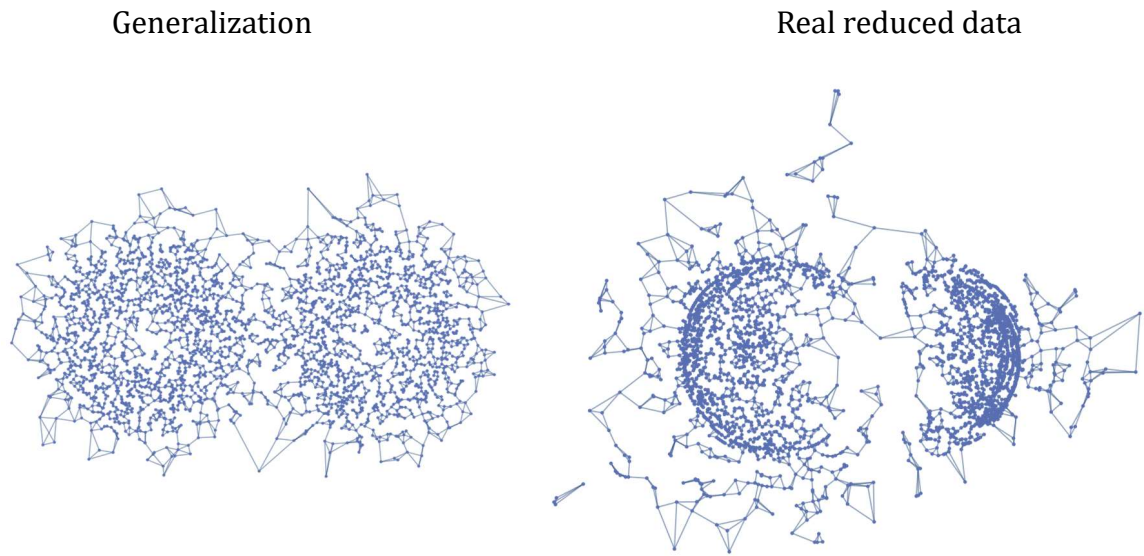

Source: Author

Looking at the real reduced data, however, one must pay attention to the displayed structure, as it resembles a single nucleus that was split in half and desires to further itself, unable to in thanks to the gatekeeping connections in the middle. This observation points to a significant feature in the real reduced data—the presence of what can be described as a central bridging mechanism within the network, which actively maintains the separation between the clusters while simultaneously preventing them from drifting further apart. This central bridging is indicative of gatekeeper profiles or key shared topics that hold influence over both groups, serving as pivotal nodes that manage the flow of discourse and maintain connectivity despite ideological differences. These gatekeepers, or central connections, play a dual role. On one hand, they ensure that there is some form of interaction between the opposing sides, which could potentially facilitate understanding or at least the coexistence of divergent views within the same network. On the other hand, they reinforce the divide by controlling the nature and extent of these interactions, ensuring that while the groups may be aware of each other and superficially connected, they remain fundamentally separate.

The role of symmetry in this context is particularly fascinating as it reveals how balanced opposing forces within the network can shape the discourse landscape. Symmetry in the network structure, where each side mirrors the other in terms of size, influence, or connectivity, suggests a form of equilibrium that can be both stabilizing, but most importantly, perpetuating. The presence of symmetry might also indicate that any disruptions or changes within one cluster could have mirrored repercussions in the opposite cluster, maintaining the overall balance of the network. For example, if one group intensifies its discourse, the other might respond similarly to maintain its standing, thereby perpetuating the polarization.

Garimella et. al. (2018) had observed this phenomena, noting how such symmetrical structures often enhance the stability and entrenchment of echo chambers. They argue that while this symmetry might superficially suggest a balanced exchange of ideas, in reality, it often exacerbates the separation between ideologically aligned groups. This is because each group, seeing its own reflections and ideologies reinforced within its chamber, might become more entrenched and less likely to engage with or even acknowledge opposing views. This reinforcement effectively solidifies the echo chamber, making it more resistant to change and more isolated from counter-narratives, which could potentially de-escalate polarization. Consequently, these symmetrical patterns in social media networks can contribute significantly to the polarization and division seen in online political discourse. They define gatekeepers as individuals who are uniquely positioned to bridge these ideologically separated groups as they have access to content from both sides of the ideological spectrum but tend to propagate content that aligns with a specific viewpoint. This selective sharing reinforces certain narratives while filtering out opposing viewpoints, thus moderating the flow of information across the echo chambers. Despite their potential to introduce diverse perspectives, Garimella et al. (2018) highlight that gatekeepers often strengthen polarization by mainly amplifying content that supports their own biases, thereby acting less as bridges and more as filters.

It seems that this role contributes to maintaining the symmetry within the network, as gatekeepers help sustain the ideological purity of the clusters they are aligned with, further deepening the divisions within the social media landscape.

**Multipolar Constellations**

Finally, we reach the Multipolar Constellations. As the name suggests, Multipolar Constellations, within the context of digital social media analysis, represent scenarios where multiple dense clusters of data points manifest, each signifying distinct areas of interaction or discussion. These clusters may represent diverse viewpoints, ideologies, or focal topics, each with its own significant concentration and influence within the same sample. The generalization may be defined as:

$$\boldsymbol{Multipolar\ Constellation} = \{\{C_1, C_2, \ldots, C_n\}, \{f_i(x,y)\}, \{D_{ij}\}\}$$

$$f_i(x,y) = exp\left(-\frac{(x-x_{cj})^2 + (y-y_{cj})^2}{2\sigma_i^2}\right)$$

$$D_{ij} = \sqrt{(x_{ci} - x_{cj})^2 + (y_{ci} - y_{cj})^2}$$

Here, the centroids $\{C_1, C_2, \ldots, C_n\}$ are the epicenters of each cluster within the constellation, analogous to the 'Nuclear' and 'Bipolar' models. In social media data, these centroids represent distinct topics, events, or influential profiles that anchor the discussion and interaction within their respective clusters. Each centroid $C_i$ is characterized by coordinates $(x_{ci}, y_{ci})$, defining the center of the cluster in the data space.

Each function $f_i(x,y)$ describes the distribution of data points around its respective centroid $C_i$, and the Gaussian model captures how interactions or discussions are concentrated around these focal points and how they taper off as one moves away from the center. The standard deviation $\sigma_i$ in each function represents the dispersion or spread of the data points around the centroid, indicating the intensity and reach of the engagement within the cluster. Finally, the distances $D_{ij}$ are the distances between each pair of centroids $\{C_i, C_j\}$ where $i \neq j$, indicates the spatial separation between the clusters. A larger $D_{ij}$ indicates a greater divide between clusters, suggesting minimal interaction and highlighting distinct linguistic, ideological, or topical spaces within the social media landscape. This metric is essential for identifying potential barriers or bridges between diverse groups or discussions.

Of all the Constellations here described, this was the one with least occurrences, which may indicate that political affairs might quickly polarize, or centralize around influential figures or pressing issues, suggesting that in digital Social Media environments, discussions often coalesce into simpler, more easily discernible patterns, either rallying around a central theme (Nuclear Constellations) or dividing into opposing viewpoints (Bipolar Constellations).

For this generalization, I have generated a sample with three centroids, though there could be more, reduced the sample through an Euclidean-based method and applied the kNN filtrations:

**Figure 15** – A set of 34000 coordinates generated from the Multipolar Constellation Generalization parameters and it's Euclidean-based reduction.

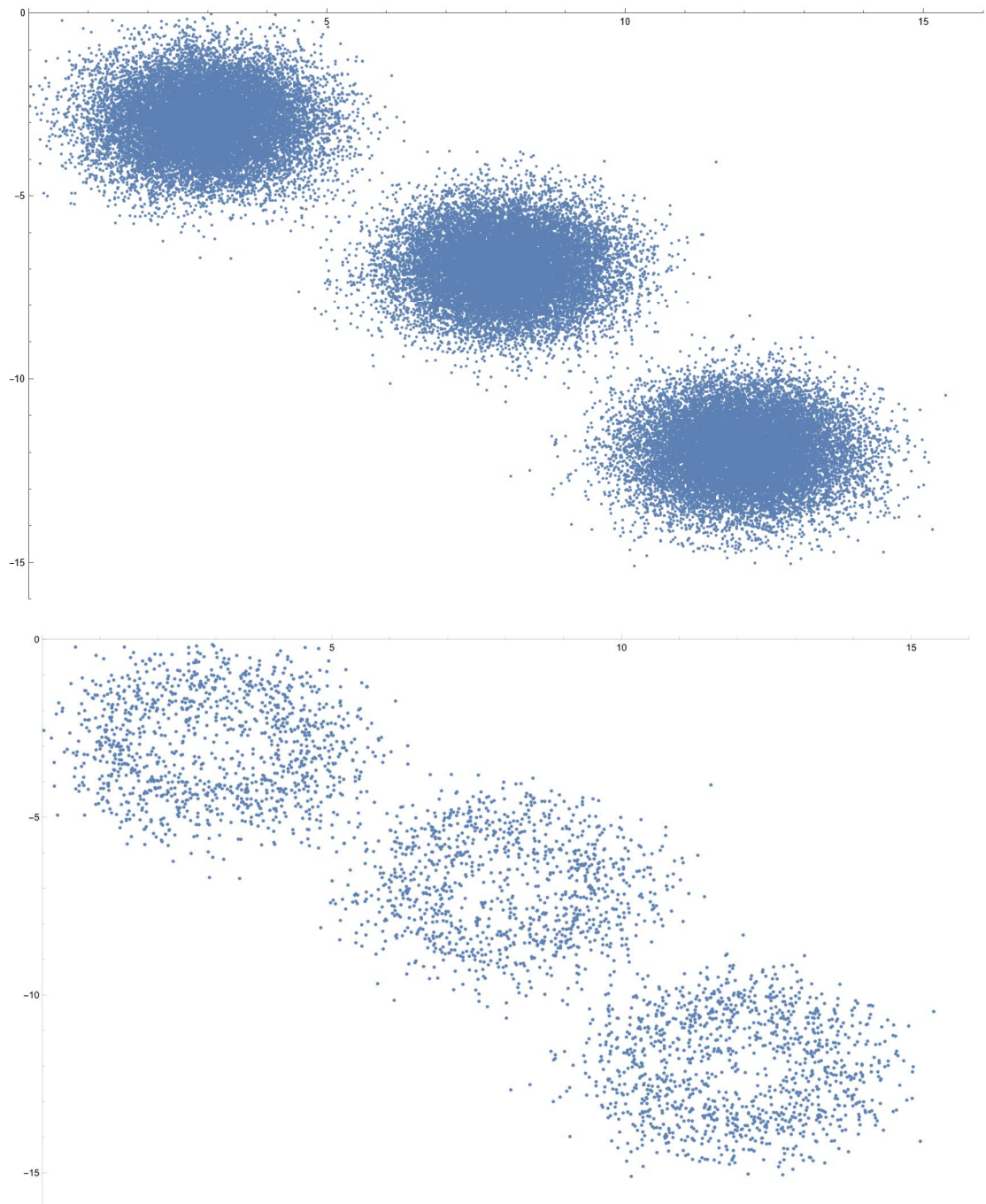

Source: Author.

**Figure 16** – kNN filtrations applied upon the reduced sample to showcase the Persistent Homology.

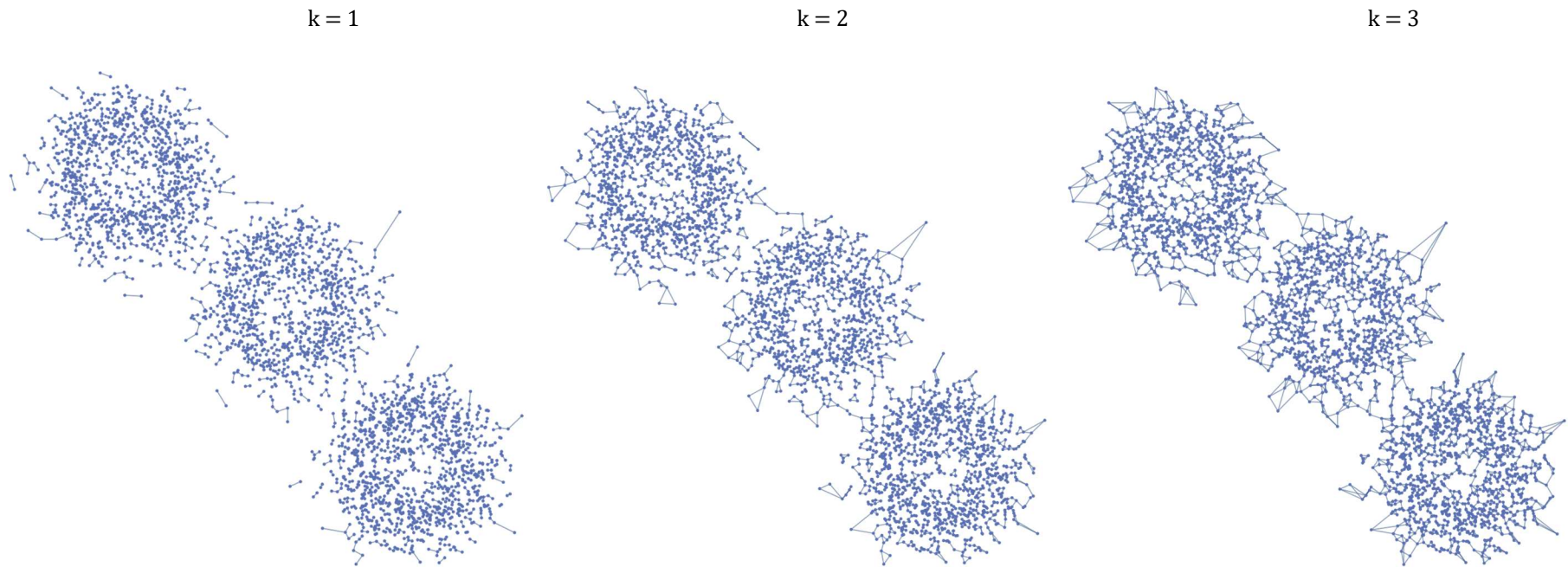

Source: Author.

*Real Data*

As previously stated, Multipolar Constellations were the least observed in the context of the data scraped throughout the 2022 Brazilian elections in Twitter. The example here showcased may be categorized as a Multipolar Constellation but, within its polarities a Bipolar Constellation may be observed, which endorses the hypothesis that overtime, a Multipolar Constellation might convert into a Bipolar Constellation. Let us, however, look into the sample drawn from the word '*Democracia*' and charted through its retweeting patterns into a network:

**Figure 17** – The Network Graph built from the retweeting user patterns of tweets containing the word '*Democracia*'.

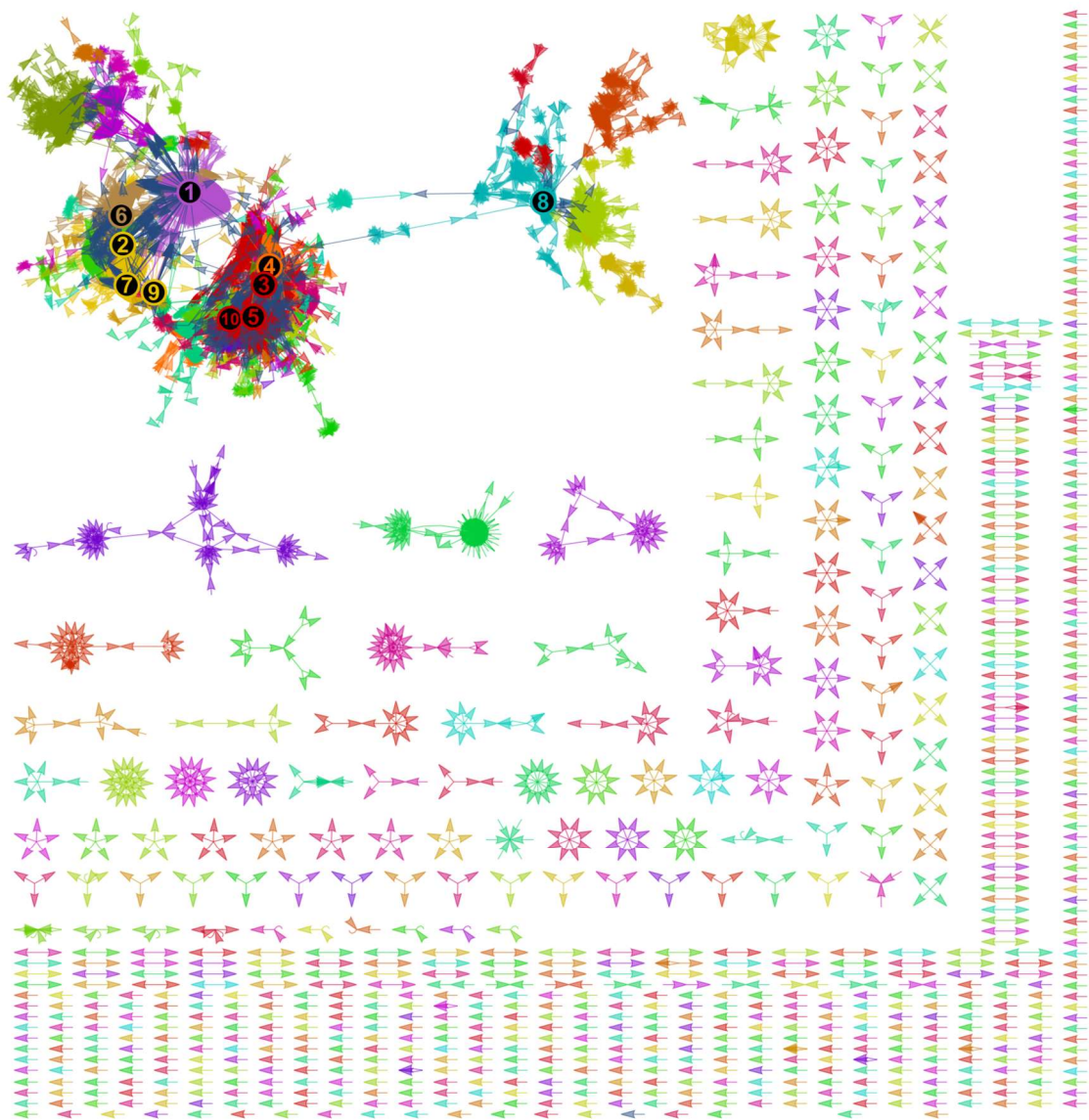

Source: Author.

As it happens, Neymar (@neymarjr), currently Brazil's most prominent football player and by all means a celebrity, is the most retweeted profile in this sample, with his medal and clique colored in purple. Now he finds himself in the midst of a heated debate regarding freedom of expression and the true nature of democracy, taunting the Left by saying that those who call themselves democratic are now attacking him by displaying an opinion that is different than their own, that is, his overall support for Bolsonaro. Bolsonaro himself (@jairbolsonaro), in yellow, is the second most retweeted profile, bolstering the narrative that Globo and the Media supports Lula and actively works for his re-election, that Globo is concerned with making more cash, and not democracy, and that during his government he decreased the public spending on publicity, which would be why he is under attack by Big Media.

Looking to the cluster on the right, populated by personalities who support Lula and the Left, a Supreme Federal Tribunal (STF)'s minister, Alexandre de Moraes, emerges as the third most retweeted profile, indicated in red. He advocates for the mottos of peace, security, respect, and harmony during the upcoming elections, considering them to be the grandest celebrations of the democratic system. Meanwhile, the fourth most retweeted profile belongs to Mark Ruffalo (@MarkRuffalo), an actor renowned for his portrayal of Marvel's Hulk and his activism in support of green policies. He shares a quote from Juliette (@juliette), a singer, lawyer, and entrepreneur, asserting that the world is closely observing Brazil's elections. Mark highlights Juliette's tweet, which urges women to exhibit bravery and exercise their right to vote. This draws attention to the unfortunate reality that many women face coercion in refraining from voting, emphasizing the importance of courageously exercising their democratic rights. The tenth most retweeted profile belongs to Antonio Tabet (@antoniotabet), an actor from *Porta dos Fundos* ("Backdoor" in English), a renowned comedy channel. Tabet asserts that honesty, penitence, and education are not commonly found in Brazil. Moreover, he strongly criticizes Bolsonaro, labeling him as the epitome of a negative role model and a significant threat to peace, security, institutions, minorities, and democracy itself.

Back to the political right, Carlos Bolsonaro (@CarlosBolsonaro) in brown and the sixth most retweeted profile of the overall sample, calls attention as to how 'people' find it all good when a global actor (Mark Ruffalo) supports 'the thief' (Lula), though find it absurd, the 'end of the world' when a football player (Neymar) supports

Bolsonaro, finding it absurd that 'people' find decades of a left-wing government all good, and four years of the right a terrible thing. Now, the seventh profile is a person profile supportive of Bolsonaro, and the 9th most retweeted profile in yellow belongs to Futmais @futtmais, a profile for football news review, delivering the message from Neto, a famous indoor soccer player, that though he does not vote for Bolsonaro, Neymar must have his right to express his political opinion.

As for the 8th profile, which breaks the cluster, something interesting happens as this cluster is not related to politics per se, but rather, the break occurs because it is a Spanish-speaking profile. And so, regardless of whether they support the political left or right, the profiles surrounding the 8th profile are speaking Spanish, creating a cluster of their own and indeed confirming what Mark Ruffalo stated: the world was watching Brazil's 2022's elections.

Let us now explore the topological features of the figure:

**Figure 18** – The coordinates selected from the *'Democracia'* main Network and it's Euclidean-based reduction.

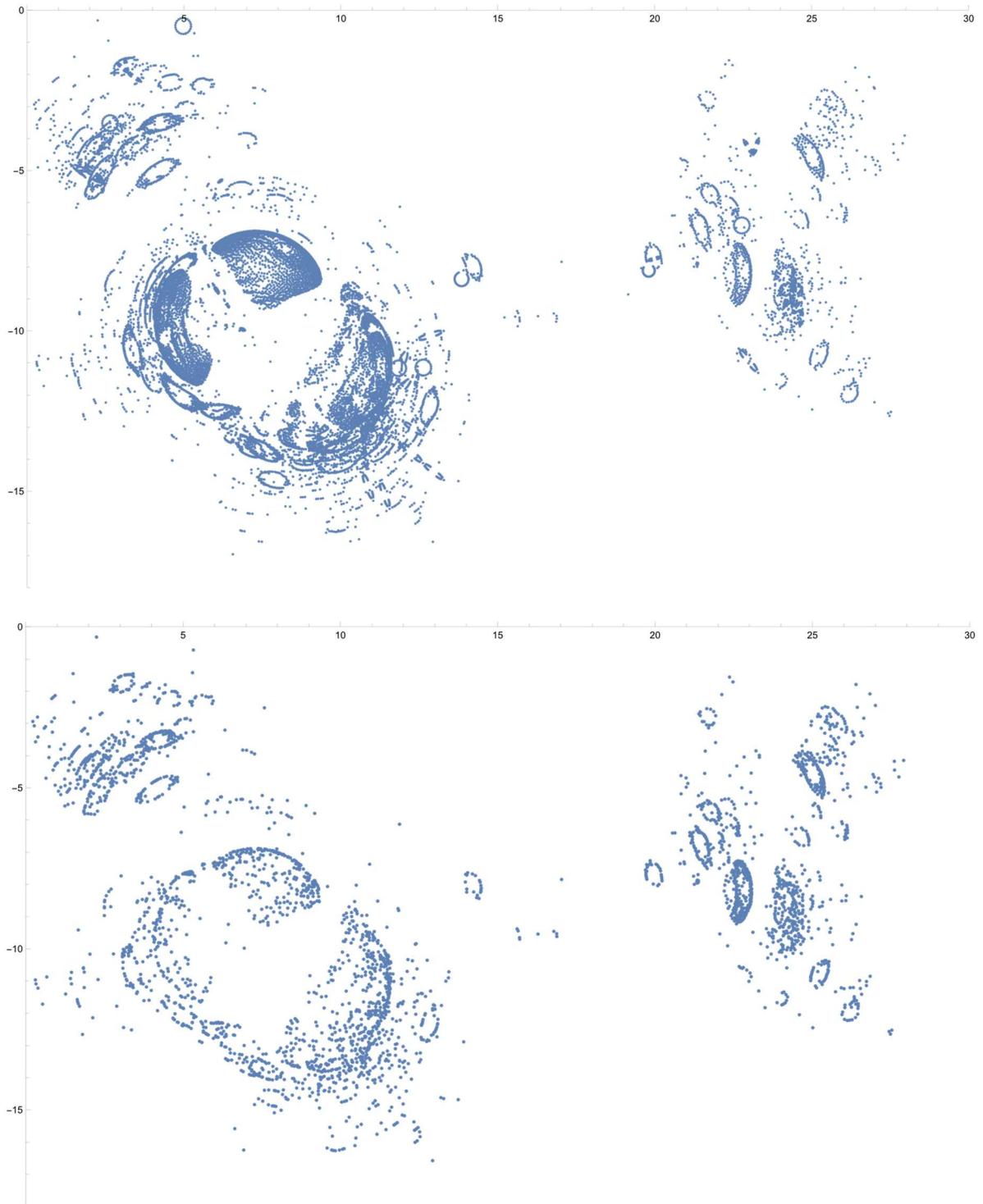

Source: Author.

**Figure 19** – kNN filtrations applied upon a reduced sample[8] of the '*Democracia*' network coordinates to showcase the Persistent Homology.

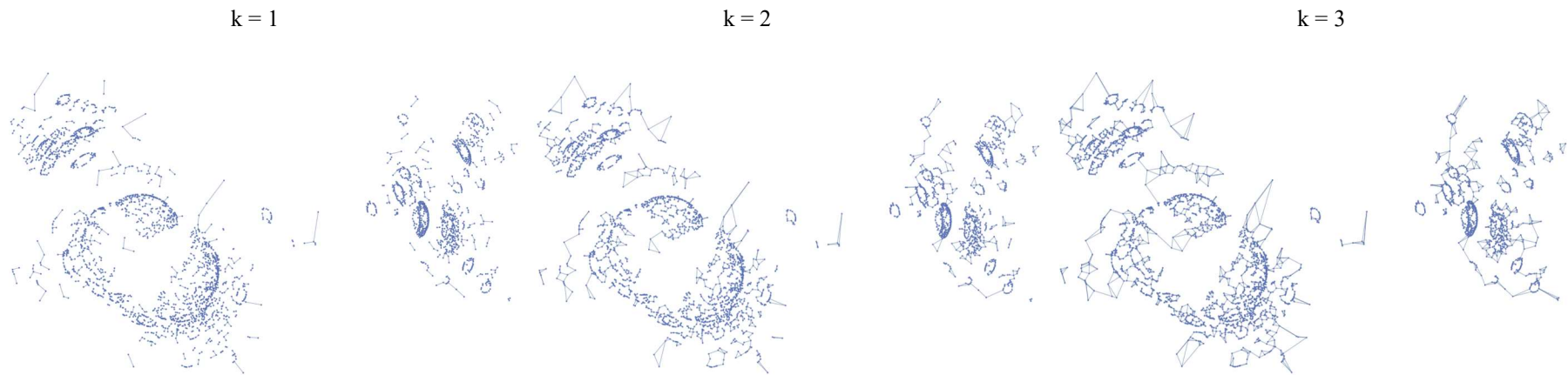

k = 1    k = 2    k = 3

Source: Author.

---

[8] The number of points used for this kNN filtration had to be reduced from the original 13919 points to 3055 to accommodate limited computation power.

The generalization model of Multipolar Constellations is defined by multiple dense clusters of data points, each with distinct ideological or topical focuses, represented mathematically as generalizations as Gaussian functions around each centroid. In the '*Democracia*' network, this is visually and quantitatively evident. For instance, the prominent clusters around key figures like Neymar and Bolsonaro represent significant ideological epicenters that draw dense interactions. These clusters are marked by their own unique narratives and discourse styles, which are distinctly visible in the network graph. Neymar's comments spark debates on the nature of democracy and freedom of expression, aligning with pro-Bolsonaro sentiments, whereas figures like Alexandre de Moraes and Mark Ruffalo champion democratic processes and civic engagement, resonating with anti-Bolsonaro and pro-democratic sentiment.

Comparatively, the generalization model suggests that larger distances $D_{ij}$ between centroids indicate minimal interactions and a greater divide, a hypothesis that is strongly supported by the real data. The minimal cross-engagement between clusters such as those centered around pro-Bolsonaro figures and those advocating for democratic integrity and opposition narratives underlines the stark ideological divides. This separation is critical as it not only underscores the challenges in fostering cross-ideological dialogue but also highlights the echo chamber effect prevalent in digital social media environments, where homogeneous groups predominantly interact within themselves, reinforcing existing beliefs and biases.

A few problems arise, however, when comparing the generalization model and the real data. First, the obvious issue is that the real data is much noisier than the clean, randomly generated data from the generalization parameters, where the centroids are clearer and more clustered. This is because, unlike homogeneously distributed points in the generalization, real social media data contains a variety of user types, ranging from highly active influencers to more passive participants. This variation affects the distribution patterns within each cluster, leading to asymmetric shapes that do not fit neatly into the Gaussian model prescribed by the generalization.

Another issue that becomes clearer when it comes to the Multipolar Constellations is that there seems to be a relation of symmetry in real data not showcased in the Gaussian generalization models. This is evident first in the Bipolar Constellations, but more so in the Multipolar Constellations, as the cluster emerging at the right of the figure seems to balance the holed shape to the left and its nuclear appendix on top.

**Figure 20** – Generalization sample (34000 points generated from the gaussian function, and then reduced by their Euclidean distance) and real data kNN (k = 3) comparison.

Generalization                              Real reduced data

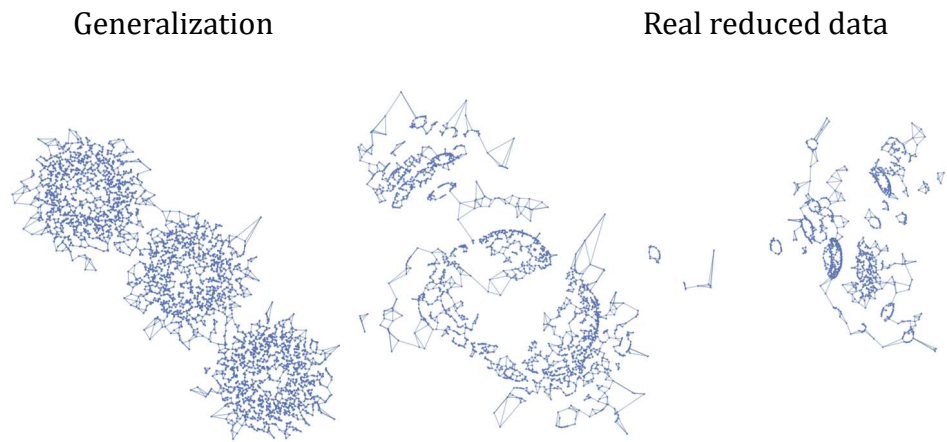

Source: Author

This differentiation only reminds us that the collected data is but a small photograph of a tiny sample of data collected over but a few hours. This implies that the network is evolutionary, and the structures we observe are not static but dynamic, shifting in response to real-time social, political, and media-driven influences. As such, the symmetry observed might reflect a momentary equilibrium or a transient phase in a broader, constantly evolving dialogue, suggesting underlying mechanisms of self-organization within the network. Social media users tend to gravitate towards content and discussions that resonate with their existing beliefs, leading to the formation of clusters. However, as opposing views interact or as users are exposed to differing perspectives, new clusters may form, and existing ones may adjust, creating a dynamic balance within the network. This self-organizing behavior is crucial for understanding how information spreads and how consensus or polarization occurs over time.

**Conclusions: the good, the bad, and the ugly**

*The Good: Gaussian Models as a stepping stone*

In this article, I have delved into the dynamics of social media through the lens of Topological Data Analysis, focusing on Persistent Homology to discern patterns of political discourse. My research has uncovered several instances of what I term as Nuclear, Bipolar, and Multipolar Constellations – each representing distinct configurations of social media interactions shaped by political narratives. The Nuclear Constellation exemplifies centralized discussions often revolving around a singular, dominant topic or personality, mirroring the phenomena of political personalism. Bipolar Constellations illustrate clear divisions within discourse, reflecting the intense polarization that characterizes much of today's political landscape. Multipolar Constellations, though less common, depict a more fragmented discourse with multiple centers of opinion, indicative of a diverse but segmented public. For the purposes of this article, I have selected an example of each constellation type to showcase their characteristics and implications comprehensively. While exploring these constellations, I found that the Gaussian model, with its emphasis on centralized peaks of activity, is most apt for describing the structure of Nuclear Constellations, where focus tends to converge sharply around central figures or themes.

The effectiveness of Gaussian functions in modeling Nuclear Constellations is rooted in their inherent ability to encapsulate the concentrated nature of discourse around singular, pivotal figures or themes that dominate the digital political arena. These functions, with their characteristic bell-shaped curves, are particularly suited to capturing how discussions or interactions intensify around a central point, diminishing in intensity as they move away from the focal topic or personality. This modeling is reflective of the centralization observed in political personalism, where a single entity or issue can command substantial attention and engagement, effectively shaping the narrative and direction of discourse within a network. Such centralized patterns are vividly illustrated in the case of political campaigns and key public figures, where strategic communication efforts often focus on core messages or personalities to rally support. The Gaussian distribution aptly represents this dynamic, highlighting how engagement peaks around these central nodes and tapers off with increasing distance from the core, both literally and metaphorically. This approach aligns with the observations in political discourse where symbolic and emotionally charged communication can resonate deeply with constituents, fostering a strong, focused community of discourse centered around charismatic leadership or pivotal issues – akin to the digital populism described by Gerbaudo (2018) as it unfolds on social media platforms. These platforms amplify such centralized, charismatic appeals, further validating the use of Gaussian models to describe the underlying structure of Nuclear Constellations within social media networks.

*The Bad: Holes and the problem of Symmetry*

One (me) must, however, realize that, despite the utility of a gaussian-based model for generalizing nuclear functions, more complexities arise as I move beyond the scope of centralized constellations into the realms of Bipolar and Multipolar configurations. In these more complex scenarios, the interplay between multiple centers of discourse and the cross-influences they exert on one another introduce dynamics that a simple Gaussian function might not adequately capture. This is particularly evident in Bipolar Constellations, where two dominant ideological poles generate separate but equally intense centers of activity, with interactions that are not merely diminishing from a central point but are dynamically interactive across a divisive spectrum. Would a Gaussian model with two centroids better suit bipolar Constellations?

In this case,

$$f(x,y) = A \exp\left(-\left(\frac{(x-\mu_1)^2}{2\sigma_x^2} + \frac{(y-\mu_1)^2}{2\sigma_y^2}\right)\right)$$

$$\mu_x = \frac{x_1 + x_2}{2} \text{ and } \mu_y = \frac{y_1 + y_2}{2}$$

$\mu_x$ and $\mu_y$ would be the coordinates of a point that averages the locations of the two centroids, reflecting a shared influence zone, and $\sigma_x$ and $\sigma_y$ with *A* representing the amplitude, possibly adjusted to reflect the intensity of the discourse around both poles. Such model would suggest that the discourse is not merely centered around each pole separately but is influenced by a kind of tug-of-war where the midpoint between the poles could become a hotbed of interaction, or potentially, a contentious battleground.

Such model would work wondrously with the reduced samples, while the first seems more suitable with larger datasets as expressed by the point cloud showcased, where the broader distribution of points could be more clearly attributed to specific ideological centers. In smaller, more focused samples where individual interactions and finer subtleties of the discourse are more pronounced, the model incorporating a single Gaussian function with dual centroids could highlight the nuanced interplay between these poles. This approach could potentially enable a deeper exploration of how ideological influence shifts across the social media landscape, particularly in scenarios where the boundaries between opposing viewpoints are blurred and the discourse is heavily contested.

This adaptation recognizes the inherently dynamic nature of social media data, which is continuously evolving as new interactions unfold and as users respond to ongoing political events. By analyzing these data snapshots, I capture only a momentary glimpse of the complex web of interactions that characterize political discourse online. Therefore, the hypothesis that a single Gaussian model with dual centroids might evolve into a model with two distinct Gaussian functions becomes particularly compelling. This transition could reflect the solidification of ideological positions over time or in response to specific events, leading to a clearer separation and identification of distinct ideological clusters. The same thing could, potentially occur on the opposite direction, with two Gaussian functions gradually merging into a single function with dual centroids if the ideological divides begin to blur or converge. Such scenarios highlight the fluid nature of social media discourse, where ideological boundaries are not always fixed but can be reshaped by emerging narratives or shifting public sentiments.

To rigorously test this hypothesis and truly understand the temporal dynamics of ideological discourse on social media, future research will need to incorporate time series analysis. This method would allow for tracking the evolution of discourse structures over time, observing how the centers of discussion emerge, converge, or diverge in response to external stimuli or internal community shifts. It would also provide insights into the persistence of certain discourse patterns and the factors that influence their stability or transformation. As these transformations occur, the models we use must be adaptable enough to accurately reflect the changing landscape of online discourse. This calls for a dynamic approach to modeling, where the parameters can adjust in response to observed changes in the data over time. This ability to track and model the evolution of discourse structures is crucial for understanding how political ideologies spread, interact, and potentially integrate within the social media environment.

Furthermore, in Multipolar Constellations, where numerous focal points emerge, the landscape becomes even more intricate. Here, the discourse does not revolve around a single or dual epicenter but is distributed among multiple centers, each with varying degrees of influence and engagement. This diversity in engagement and influence challenges the Gaussian model's ability to represent the spread and interaction of discourse effectively. The interconnections and the subtle nuances of influence and exchange among these multiple centers require a more flexible and multidimensional analytical approach, perhaps incorporating a combination of Gaussian functions or exploring entirely different mathematical models that can better handle the complexity and dynamism of such configurations.

Building on the insights gleaned from the preliminary exploration of Gaussian models with dual centroids, further research could fruitfully investigate the role of symmetry in these models. The hypothesis that symmetry within the model could either facilitate a more balanced discourse or entrench existing polarizations offers a rich avenue for exploration. For instance, the research question, "What role does symmetry play in this model?" seeks to uncover whether symmetric distributions of discourse centers lead to more stabilized or more fragmented ideological landscapes. Additionally, exploring whether symmetry can be employed to analyze polarization, encapsulated in the question "Can symmetry be employed to analyze polarization?", will help determine if symmetric properties of the model correlate with greater or lesser degrees of ideological divide within the discourse.

Another compelling area of investigation involves the mathematical exploration of gatekeepers within these models. Gatekeepers, as identified in previous discussions, serve pivotal roles in modulating the flow of information between disparate ideological groups. The research question, "How can we explore the role of gatekeepers mathematically?" aims to develop quantitative measures or indicators within the Gaussian model framework that can identify and assess the impact of gatekeepers. This could involve adjusting the model to account for nodes that disproportionately influence the transfer of discourse between clusters or analyzing the gatekeepers' positions within the symmetry of the ideological landscape.

To address these questions, advanced computational techniques and methodologies will be required. Incorporating machine learning algorithms to detect patterns of symmetry and the role of gatekeepers, or employing network analysis tools to visualize and quantify the structural dynamics of discourse, could provide deeper insights. Moreover, employing time series analysis as part of this broader research agenda will be essential. This approach will allow researchers to observe the temporal shifts in the discourse structures, providing a dynamic view of how political narratives evolve and how gatekeepers and symmetry contribute to the persistence or transformation of these narratives over time.

### *The Ugly: Hardware limitations and further research*

Therefore, the evolutionary nature of these networks highlights the need for adaptive analytical models that can account for the fluidity and complexity of social media landscapes. Persistent Homology is a powerful tool in topological data analysis that can reveal hidden structures within data that other methods might overlook. However, its application in social media analysis faces significant challenges, primarily due to hardware limitations that necessitate data reduction. This reduction can sometimes obscure the nuanced interactions and structures that are critical for understanding complex social media dynamics. The point clouds Point clouds generated from social media networks provide a visually intuitive method to apply Gaussian models or the dual-centroid Gaussian model I suggested. These models, particularly when used with larger datasets as indicated by the point clouds, can more effectively attribute ideological shifts and concentrations within the data. However, the necessity to reduce datasets for Persistent Homology analysis means that some of the finer topological features were lost, making it difficult to accurately capture the true complexity of the discourse.

To overcome these limitations and harness the full potential of Persistent Homology for large datasets, an advanced computational approach is needed. One possible solution could involve the use of distributed computing or parallel processing techniques, which can handle larger datasets more efficiently by dividing the workload across multiple processors. Additionally, the development of more efficient algorithms for Persistent Homology that require less computational power could also be a crucial step forward. Researchers more proficient in coding could focus on algorithm optimization specifically tailored for social media data, potentially reducing the complexity of computations or the amount of data needed to detect significant topological features.

Another ugly discovery was that some Constellations seem to be found in the intersection of Bipolar and Nuclear constellations – displaying holes! These holes, which are critical in TDA for identifying and understanding the underlying structures of data, reveal gaps or voids within the discourse networks, possibly indicating missing links or isolated subgroups within the broader conversation – or outright aversion of a groups within a single discussion. This intersection and the resultant topological features highlight the complexity of political discourse on social media, where not all interactions are straightforward or easily classified into distinct categories.

The presence of holes in the data might suggest a snapshot of a moment, capturing the fluidity and transient nature of social media discourse. These topological gaps could represent the emerging or dying of certain topics or the shifting allegiances and fragmentations within ideological groups. Such insights are invaluable as they provide a deeper understanding of the dynamics at play within digital communities. However, fully exploring these topological features and their implications requires computational resources beyond what is typically available for standard analyses. Only a more powerful computational tool could allow me to effectively explore these holes and other intricate topological figures inherent in the data. This would involve leveraging high-performance computing or advanced data processing technologies that can handle the intensive calculations required by TDA without compromising the integrity of the dataset.

An hypothesis arising from these observations is that these topological figures displaying holes might not only suggest a snapshot of the moment but also showcase the inherent fluidity and ever-changing nature of discourse within social media. They indicate areas where the discourse is not just divided but fundamentally disconnected, potentially highlighting critical points for intervention or further study to understand what drives these disconnections. Nevertheless, not enough topologies with holes were discovered to hypothesize a relation between holed-topologies and political procecesses.

Lastly, the interaction between clusters in real data can be affected by the platform's algorithms, which can either promote or suppress certain types of content, thereby influencing the flow of discourse. This algorithmic mediation can alter the natural interaction patterns that would otherwise be observed, adding an additional layer of complexity to the analysis of digital social media landscapes using the Multipolar Constellation model. An important research question that arises from this observation is, "How does the model work on other social media platforms?" This question aims to explore whether the dynamics observed on one platform hold consistent across others, which may have different user demographics, content promotion algorithms, and interaction mechanisms. Understanding these differences is crucial for developing a more universally applicable model that accurately captures the nuances of political discourse across various digital environments.

Bad and ugly seems more than just good, but, overall, it is not so much of a terrible thing when research generate more problems and questions. In fact, the emergence of new questions and challenges is a hallmark of progressive inquiry, pushing the boundaries of understanding and prompting deeper exploration. Each new question or problem uncovered during research acts as a stepping stone toward greater knowledge, guiding further studies and innovations. It encourages a continuous cycle of refinement and development within the scientific and academic communities, much needed specially in the newfound realm of computational social sciences. By identifying and confronting the bad and the ugly, researchers can forge paths to new solutions and insights, ultimately contributing to a more nuanced and sophisticated grasp of digital communication dynamics. Moreover, the complications and complexities encountered along the way enrich the researcher's toolkit, fostering a more robust approach to tackling similar issues in the future.

Therefore, while it may seem disheartening at first to encounter more questions than answers, I can only hope that this phenomenon is actually indicative of thorough and effective research, and have faith that, in the future, this research will be beautiful, worse and hideous.